\documentclass[final]{ias2}
\pdfoutput=1

\usepackage{graphicx} 
\usepackage{multirow}
\usepackage{array} 
\usepackage{latexsym}
\usepackage{amssymb}

\ifx\pdfoutput\undefined
\usepackage[dvips,bookmarks]{hyperref}	% This is for arXiv.org
\else
\usepackage{hyperref}	% This is for pdftex
\fi
\hypersetup{colorlinks,bookmarksopen,bookmarksnumbered,citecolor=verdes,
linkcolor=blus,pdfstartview=FitH,urlcolor=rossos}
\def\myurl#1#2{\href{http://#1}{#2}}
\def\hhref#1{\href{http://arxiv.org/abs/#1}{#1}} % in bibliography
\usepackage{color}
\definecolor{rosso}{cmyk}{0,1,1,0.4}
\definecolor{rossos}{cmyk}{0,1,1,0.55}
\definecolor{rossoc}{cmyk}{0,1,1,0.2}
\definecolor{blu}{cmyk}{1,1,0,0.3}
\definecolor{blus}{cmyk}{1,1,0,0.6}
\definecolor{bluc}{cmyk}{1,1,0,0.1}
\definecolor{verde}{cmyk}{0.92,0,0.59,0.25}
\definecolor{verdec}{cmyk}{0.92,0,0.59,0.15}
\definecolor{verdes}{cmyk}{0.92,0,0.59,0.4}

\begin{document}

\markboth{Indirect Searches for Dark Matter}{Cirelli Marco}

\title{Indirect Searches for Dark Matter: a status review}

\author[cern,ipht]{Marco Cirelli}
\email{marco.cirelli@cea.fr, marco.cirelli@cern.ch}
\address[cern]{CERN Theory Division,  CH-1211 Gen\`eve, Switzerland}
\address[ipht]{IPhT, CNRS, URA 2306 \& CEA/Saclay, F-91191 Gif-sur-Yvette, France}

\begin{abstract}
I review in a schematic way the current status of indirect searches for Dark Matter: I list the main relevant experimental results of the recent years and I discuss the excitements and disappointments that their phenomenological interpretations in terms of almost-standard annihilating Dark Matter have brought along. I then try to individuate the main directions which have emerged from the recent very intense model-building activity. In passing, I list the main sources of uncertainties that affect this kind of searches.\\ {\em [Report number: Saclay T11/206, CERN-PH-TH/2011-257]}\\ 
{\em [Prepared for the Proceedings of Lepton-Photon 2011, Mumbai, India, 22-27 Aug 2011]}
\end{abstract}


 
\maketitle

% \tableofcontents
% \listoffigures
% \listoftables

\section{Introduction}
\label{introduction}

Cosmology and astrophysics provide several convincing {\bf evidences of the existence of Dark Matter} (DM). 
The observation that some mass is missing to explain the internal dynamics of galaxy clusters and the rotations of galaxies dates back respectively to the '30s and the '70s. The observations from weak lensing, for instance in the spectacular case of the so-called `bullet cluster', provide evidence that there is mass where nothing is optically seen. More generally, global fits to a number of cosmological datasets (Cosmic Microwave Background, Large Scale Structure and also Type Ia Supernovae) allow to determine very precisely the amount of DM in the global energy-matter content of the Universe at $\Omega_{\rm DM} h^2 =0.1123 \pm 0.0035$~\cite{cosmoDM}\footnote{Here $\Omega_{\rm DM} = \rho_{\rm DM}/\rho_c$ is defined as usual as the energy density in Dark Matter with respect to the critical energy density of the Universe $\rho_c = 3 H_0^2/8\pi G_N$, where $H_0$ is the present Hubble parameter. $h$ is its reduced value $h = H_0 / 100\ {\rm km}\, {\rm s}^{-1} {\rm Mpc}^{-1}$.}.
 
All these signals pertain to the gravitational effects of Dark Matter at the cosmological and extragalactical scale. Searches for explicit manifestation of the DM particles that are supposed to constitute the halo of our own galaxy (and the large scale structures beyond it) have instead so far been giving negative results, but this might be on the point of changing. 

\medskip

{\bf Indirect searches} for Dark Matter aim at detecting the signatures of the annihilations or decays of DM particles in the fluxes of Cosmic Rays (CRs), intended in a broad sense: charged particles (electrons and positrons, protons and antiprotons, deuterium and antideuterium), photons (gamma rays, X-rays, synchrotron radiation), neutrinos. 
In general, a key point of all these searches is to look for channels and ranges of energy where it is possible to beat the background from ordinary astrophysical processes. This is for instance the basic reason why searches for charged particles focus on fluxes of antiparticles (positrons, antiprotons, antideuterons), much less abundant in the Universe than the corresponding particles, and searches for photons or neutrinos have to look at areas where the DM-signal to astro-noise ratio can be maximized. 

Pioneering works have explored indirect detection (ID) as a promising avenue of discovery since the late-70's. Since then, innumerable papers have explored the predicted signatures of countless particle physics DM models. In the past 3 years or so, however, the field has experienced a significant burst of activity, mainly due to the results presented by a few very well performing experiments, above all the PAMELA satellite, the FERMI satellite and the HESS telescope. It is fair to say that the field has passed, for better or for worse, from a theory-driven state to a data-driven phase. 

\medskip

The {\bf scope of this work} is to schematically present the {\em status} of the field of indirect DM detection~\footnote{For a similar effort, see~\cite{SerpicoTAUP2011}.}, with a specific attention to the experimental results and their phenomenological interpretation in terms of Dark Matter (in Sec.~\ref{status}) and with some attention to future expectations. What this write-up does not intend to be is a technical review of the methods and formul\ae\ employed for DM ID (which can instead be found to a large extent in~\cite{Cirelli:2010xx}): indeed, there is hardly any equation in this work. It also does {\em not} intend to be a proper, comprehensive backward-looking review of the activity mentioned above. Nevertheless, an attempt to isolate the main theory directions which have emerged, and the bits which are likely to stay with us, is made in Sec.~\ref{theory}. In Sec.~\ref{uncertainties} I list the main uncertainties that affect the interpretation of astrophysical signals in terms of Dark Matter: this is an important area for the DM practitioners to follow, because major advancements in the field of indirect detection will probably pass through related advancements there. Before moving to the subject matter, let us quickly remind ourselves of the framework inside which most of the activity develops: the one centered around WIMPs.

\subsection{Under the WIMP spell, or only slightly outside}
\label{wimp}

A well spread theoretical prejudice wants the DM particles to be {\bf thermal relics} from the Early Universe. They were as abundant as photons in the beginning, being freely created and destructed in pairs when the temperature of the hot plasma was larger then their mass. Their relative number density started then being suppressed as annihilations proceeded but the temperature dropped below their mass, due to the cooling of the Universe. Finally the annihilation processes also froze out as the Universe expanded further. The remaining, diluted abundance of stable particles constitutes the DM today. As it turns out, particles with weak scale mass ($\sim 100\, {\rm GeV} - 1\, {\rm TeV}$) and weak interactions could play the above story remarkably well, and their final abundance would automatically (miracolously?) be the observed $\Omega_{\rm DM}$. While this is certainly not the only possibility, the mechanism is appealing enough that a several-GeV-to-some-TeV scale DM particle with weak interactions ({\bf WIMP}) is often considered as the most likely DM candidate.

{\bf Variations} of this paradigm are of course possible, and have actually started to gain strength in recent years, as the experimental program to search for WIMPs reaches culmination. A close relative of the WIMP paradigm is the class of so-called WIMPless models, in which the thermal freeze-out story still is at play but the role of weak interactions is filled by some other kind of interaction. Indeed, very close to this class fall all the models postulating `dark forces', `secluded DM' and the like, which will be briefly addressed in Sec.~\ref{theory}. The somewhat opposite routes, instead, are to assume a different cosmological history~\cite{Gelmini:2010zh}, or to consider very feebly interacting particles which would have never reached thermal equilibrium \cite{Boyarsky:2009ix,Hall:2009bx}, or to postulate that DM originates from the decay of some other species, the latter maybe being itself a WIMP~\cite{Feng:2010tg}. Finally, a recently-popular-again suggestion is that DM is produced with a primordial asymmetry between particles and antiparticles, similarly to baryons, and that only one of the two species (say, particles, for definiteness) survives, exactly like for baryons~\cite{aDM}. In this case, there is little point in speaking of indirect search, since annihilations of DM particles are not possible for lack of target antipartices. Unless a mechanism such as DM-antiDM oscillations  intervenes to re-equilibrate the populations at late times and therefore re-switches on DM ID signals~\cite{Cirelli:2011ac,Tulin:2012re}.

Another belief which adds motivation to associating DM with the TeV scale is that {\bf New Physics} is expected to show up at that scale, essentially to cure the Standard Model hierarchy problem. Such New Physics, whatever form it takes (SuperSymmetry, ExtraDimensions, compositeness...) likely encompasses a number of new particles, among which (the conviction is) the one constituting the Dark Matter. 

\smallskip

In any case, even independently of the theory prejudices, this mass range ({\bf TeV-ish DM}) has the best chances of being thoroughly explored in the near future by charged particle and photon observatories, also in combination with direct DM searches (aiming at detecting the nuclear recoil produced by a passing DM particle in ultra-low background underground detectors) and, possibily, production at CERN's Large Hadron Collider. With notable exceptions (axions, KeV sterile neutrinos...) which however I will not discuss further, the TeV-ish ballpark is therefore the focus of the attention of the largest majority of the DM ID community.

An important corollary of the long-term fascination of the community for the WIMP miracle, or more generally the thermal relic production mechanism, is that DM particles are expected to {\bf annihilate in pairs} into Standard Model particles. More precisely, a velocity averaged annihilation cross section of $\langle \sigma v \rangle = 3 \cdot 10^{-26} \ {\rm cm}^3/{\rm s}$ is seen as the benchmark value, since it is the one that yields the correct relic abundance. Deviations from this scheme are of course possible and have actually already been mentioned. For instance, DM that decays. Or that annihilates into non-SM new states. \\
In this write-up I pay the due respect to the long-term fascination, and also to the historical development: possible hints and bounds of DM in the cosmic rays are interpreted in the frame of DM particles {\em annihilating into pairs of SM particles}. Then, in Sec.~\ref{theory}, I will briefly address the deviations from this scheme.

%%%%%%%%%%%%%%%%%%%%%%%%%%%%%%%%%%%%%%%%%%%%%%%
\section{Status of the searches and of the interpretation in terms of Dark Matter}
\label{status}

\subsection{Charged cosmic rays}
\label{charged}

\begin{figure}[t]
\begin{center}
\includegraphics[width=0.32\columnwidth]{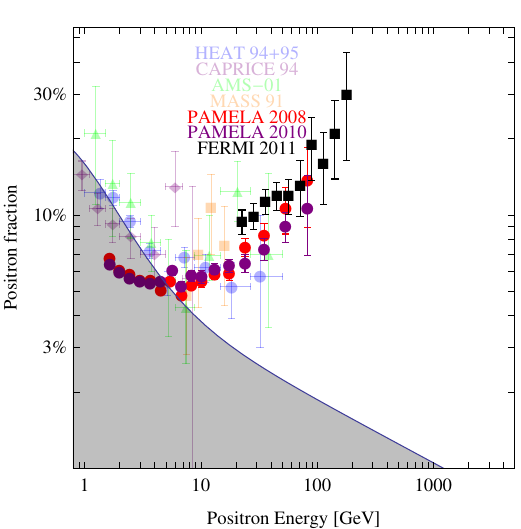} \
\includegraphics[width=0.32\columnwidth]{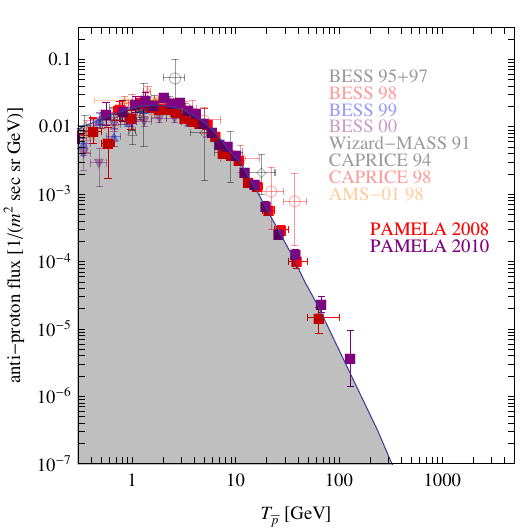} \
\includegraphics[width=0.32\columnwidth]{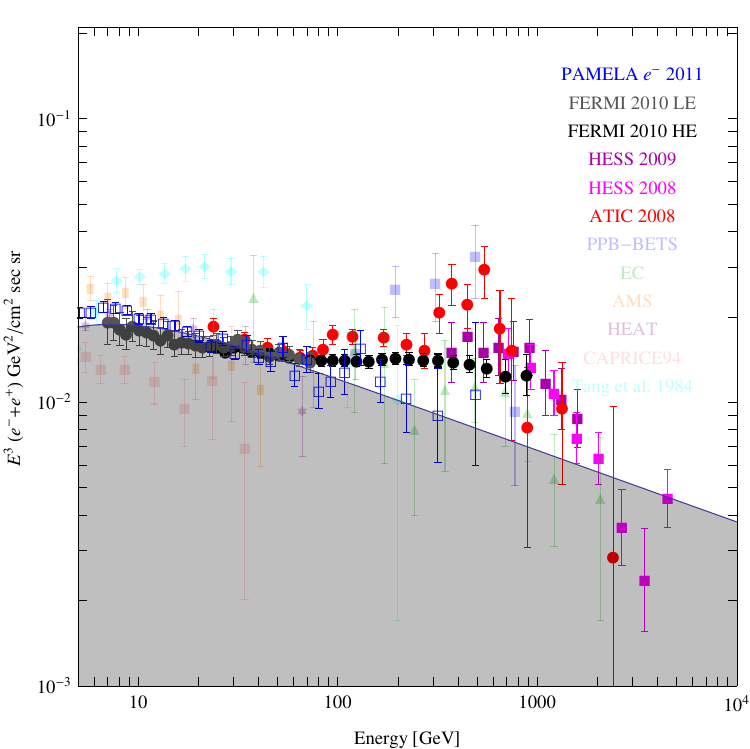}
\caption{A compilation of recent and less recent data in charged cosmic rays, superimposed on plausible but uncertain astrophysical backgrounds from secondary production. Left: positron fraction. Center: antiproton flux. Right: sum of electrons and positrons.}
\label{fig:chargeddata}
\end{center}
\end{figure}

There has been a flurry of positive results from a few indirect detection experiments looking at the fluxes of charged cosmic rays.
In particular, the signals pointed to an {\bf excess of electrons and positrons} at the TeV and sub-TeV scale: 
\begin{itemize}
\item[$\circ$] Data from the PAMELA satellite~\cite{PAMELApositrons} showed a steep increase in the energy spectrum of the positron fraction $e^+/(e^++e^-)$ above  10~GeV up to 100~GeV, compatibly with previous hints from HEAT~\cite{HEAT} and AMS-01~\cite{AMS-01}.
\item[$\circ$] Recently, these findings have been confirmed with an independent measurement by the FERMI satellite~\cite{FERMIpos}, and extended to about 200 GeV.
\item[$\circ$] Data from PAMELA~\cite{PAMELApbar} also showed no excess in the $\bar p/p$ energy spectrum compared with the predicted background.
\item[$\circ$] The balloon experiments ATIC-2~\cite{ATIC2} and PPB-BETS~\cite{PPB-BETS} were reporting the presence of a peak in the $e^++e^-$ energy spectrum at around  500-800~GeV.
\item[$\circ$] This sharp feature has been later questioned and superseded by the results of the FERMI satellite~\cite{FERMIleptons}: while an excess with respect to the expected background is confirmed, the $e^++e^-$ spectrum is found to be instead reproduced by a simple power law.
\item[$\circ$] The HESS telescope also reports the measurement of the $e^++e^-$ energy spectrum above energies of 600 GeV~\cite{HESSleptons}, showing a power law spectrum in agreement with the one from FERMI and eventually a steepening at energies of a few TeV.
\end{itemize}
The data are displayed in fig.~\ref{fig:chargeddata}, together with the expected {\bf astrophysical `backgrounds'}. The latter ones are uncertain and are an interesting subject of study by themselves in CR physics. For instance, the background positrons are thought to originate as byproducts (`secondaries') of the spallations of other CRs on the interstellar medium, but the precise prediction of their spectral slope and overall normalization is far from easy. In this vein, indeed, there have been initial suggestions attempting to `explain away' (part of) the PAMELA rise in terms of modified secondary spectra~\cite{Delahaye:2008ua}, e.g. with a dip in the $e^-$ flux which enters in the denominator of the positron fraction. However, on the basis of pretty general CR propagation arguments and also in the light of subsequent measurements of the pure $e^-$ flux by PAMELA and FERMI, these kinds of explanations have lost strenght~\cite{Serpico:2008te,Serpico:2011wg}.
 
The signals presented above are therefore striking because they imply the existence of a {\bf source of {\em `primary'} $e^+$ (and $e^-$)} other than the ordinary astrophysical ones. This unknown new source can well be itself of astrophysical nature~\footnote{...and it would actually be one of the wisest conclusions, in the light of all the rest discussed in this paper.}, e.g. one or more pulsar(s) / pulsar wind nebula(\ae), supernova remnants etc~\cite{Serpico:2011wg}. 
It is however very tempting to try and read in these `excesses' the signature of DM.

\subsubsection{Scent of Dark Matter}
\label{chargedscent}

\begin{figure}[t]
\begin{center}
\includegraphics[width=0.32\columnwidth]{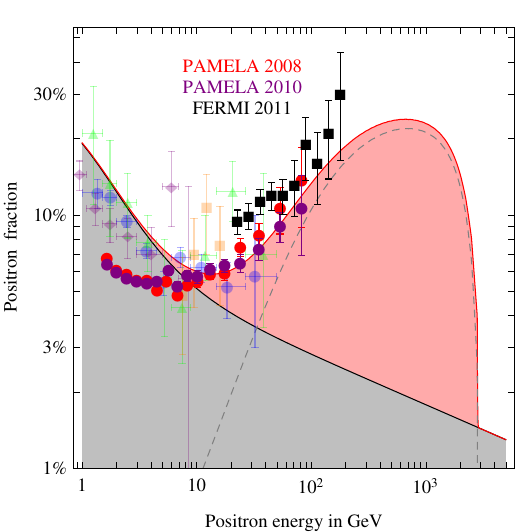} \
\includegraphics[width=0.32\columnwidth]{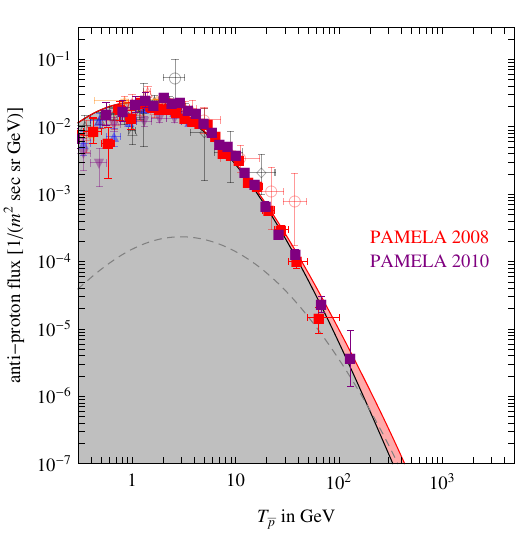} \
\includegraphics[width=0.32\columnwidth]{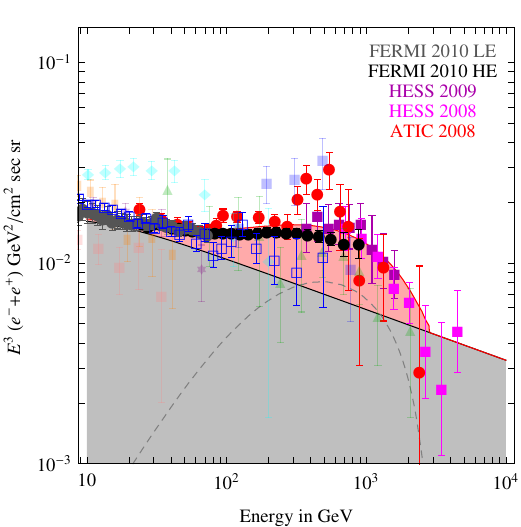}
\caption{Charged cosmic ray data interpreted in terms of Dark Matter annihilations: the flux from the best fit DM candidate (a 3 TeV DM particle annihilating into $\tau^+ \tau^-$ with a cross section of $2 \cdot 10^{-22}\, {\rm cm}^3/{\rm sec}$) is the lower dashed line and is summed to the supposed background, giving the pink flux which fits the data. Left, center and right like in fig.~\ref{fig:chargeddata}.}
\label{fig:chargedfit}
\end{center}
\end{figure}

As already mentioned above, the DM particles that constitute the DM halo of the Milky Way are expected to annihilate into pairs of primary SM particles (such as $b \bar b$, $\mu^+\mu^-$, $\tau^+\tau^-$, $W^+W^-$ and so on) which, after decaying and through the processes of showering and hadronizing, give origin to {\bf fluxes of energetic cosmic rays: $e^-, e^+, \bar p$} (and also $\gamma$-rays, $\nu$...), denoted $dN_{f}/dE$.  Depending on which one has been the primary SM particle, the resulting spectra differ substantially in the details. Generically, however, they feature a `bump'-like shape, characterized by a high-energy cutoff at the DM particle mass and, for $e^\pm$ in particular, a softly decreasing tail at lower energies (see e.g. the examples in fig.~\ref{fig:chargedfit}). It is thus clear that it is very natural to expect a DM source to `kick in' on top of the secondary background and explain the $e^\pm$ excesses. The energy range, in particular, is tantalizingly right: the theoretically preferred TeV-ish DM would naturally give origin to TeV and sub-TeV bumps and rises.

\bigskip

The $e^-$, $e^+$ and $\bar p$ produced in any given point of the halo {\bf propagate} immersed in the turbulent galactic magnetic field. The field consists of random inhomogeneities that act as scattering centers for charged particles, so that their journey can effectively be described as a diffusion process from an extended source (the DM halo) to some final given point (the location of the Earth, in the case of interest). While diffusing, charged CRs experience several other processes, and in particular energy losses due to synchrotron radiation, Inverse Compton Scattering (ICS) on the low energy photons of the CMB and starlight, Coulomb losses, bremsstrahlung, nuclear spallations...\,.
Quantitatively, the steady-state number density $n_f(\vec x,E)$ per unit energy $E$  of the cosmic ray species $f$ $(= e^+,e^-, \bar p$) in any given point $\vec x$ obeys to a diffusion-loss equation~\cite{SalatiCargese} 
\begin{eqnarray}
\label{eq:diffeq}
-\mathcal{K}(E) \cdot \nabla^2 n_f - \frac{\partial}{\partial E}\left( b(E,\vec x) \, n_f \right) &+& \frac{\partial}{\partial z}\left( {\rm sign}(z)\, V_{\rm conv} \, n_f \right) \nonumber \\
&=& Q(E,\vec x) -2h\, \delta(z)\, \Gamma \, n_f\,.
\end{eqnarray}
The first term accounts for diffusion, with a coefficient conventionally parameterized as $\mathcal{K}(E)=\mathcal{K}_0 (E/{\rm GeV})^\delta$. The second term describes energy losses: the coefficient $b$ is position-dependent since the intensity of the magnetic field (which determines losses due to synchrotron radiation) and the distribution of the photon field (which determines losses due to ICS) vary across the galactic halo.
The third term deals with convection while the last term accounts for nuclear spallations, that occur with rate $\Gamma$ in the disk of thickness $h \simeq 100$ pc.
The different processes described above have a different importance depending on the particle species: the journey of electrons and positrons is primarily affected  by synchrotron radiation and inverse Compton energy losses, while for antiprotons these losses are negligible and convection and spallation dominate. \\
The source, DM annihilations, is given by $Q = 1/2 \left(\rho(\vec x)/m_{\mbox{\tiny DM}}\right)^2 \sum_{i} {\rm BR}_i \langle \sigma v\rangle $ $(dN_f^i/dE)$, where $m_{\mbox{\tiny DM}}$ is the DM mass, $\sigma v$ is the total annihilation cross section and the sum runs over all primary channels $i$ in which the cosmic ray species $f$ is produced. $\rho(\vec x)$ is the {\bf DM density distribution} in the galactic halo. What to adopt for the latter is another one of the main open problems in the field. Based on the results of increasingly more refined numerical simulations or on direct observations, profiles that differ even by several orders of magnitude at the Galactic Center are routinely adopted: e.g. the classical Navarro-Frenk-White (NFW) or the Einasto one, which exhibit a cusp at the galactic center, or the truncated isothermal or the Burkert one, which feature a central core. All profiles, on the other hand, are roughly normalized at the same value at the location of the Earth ($\approx$ 0.3 GeV/cm$^3$). These features generically imply that observables which depend mostly on the local DM density (for instance, the flux of high energy positrons, which cannot come from far away due to energy losses) will not be very affected by the choice of profile, while those that are sensitive to the density at the GC will be affected the most (e.g. gamma rays observations of regions close to the GC).\\
Eq.~(\ref{eq:diffeq}) is usually solved numerically in a diffusive region with the shape of a solid flat cylinder that sandwiches the galactic plane, with height $2L$ in the $z$ direction and radius $R=20\,{\rm kpc}$ in the $r$ direction. The location of the solar system corresponds to $\vec x_\odot  = (r_{\odot}, z_{\odot}) = (8.33\, {\rm kpc}, 0)$.
Boundary conditions are imposed such that the number density $n_f$ vanishes on the surface of the cylinder, outside of which the charged cosmic rays freely propagate and escape. The values of the propagation parameters $\delta$, $K_0$, $V_{\rm conv}$ and $L$ are deduced from a variety of (ordinary) cosmic ray data and modelizations.

\medskip

The datasets listed in~\ref{charged} {\bf pin-point the properties of the DM particle} needed to interpret them in terms of annihilations quite precisely. The DM has to be: 
\begin{itemize}
\label{properties}
\item[$\triangleright$] With a mass of a {\em few TeV}, in order to reproduce the feature in the $e^++e^-$ spectrum. 
\item[$\triangleright$] {\em Leptophilic}, i.e. annihilating almost exclusively into leptonic channels, otherwise the antiproton measurements would be exceeded. 
\item[$\triangleright$] With a {\em very large annihilation cross section}, of the order of $10^{-23}\, {\rm cm}^3/{\rm sec}$ or more (for the masses under consideration), much larger than the thermal one, in order to produce a large enough flux that can fit the positron rise and the $e^++e^-$ bump. 
\end{itemize}
\begin{figure}[t]
$
\begin{array}{c}
$\includegraphics[width=0.4\textwidth]{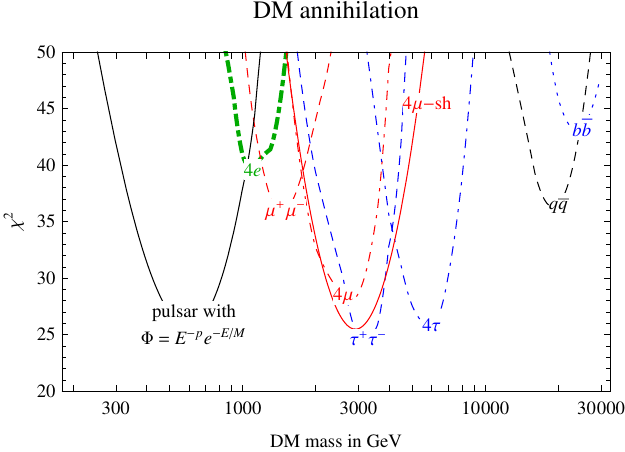}$ \
$\includegraphics[width=0.45\textwidth]{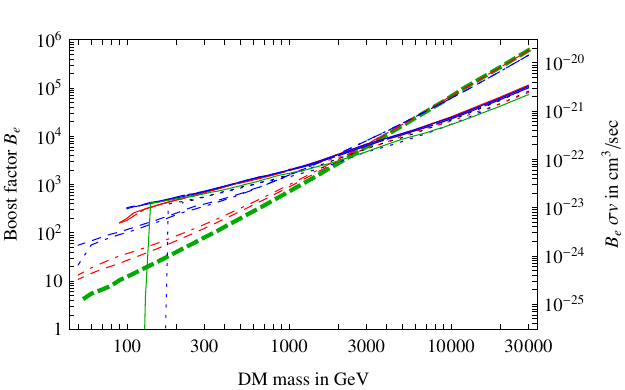} $
\raisebox{5mm}{$\includegraphics[width=0.09\textwidth]{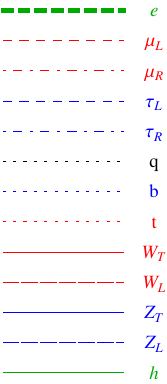}$}
\end{array}
$
\caption{Left: Global fit of different DM annihilation channels to the PAMELA, FERMI and HESS data.
The labels on each curve indicate the primary annihilation channel (figure from~\cite{Meade:2009iu} (2009); the fit results remain essentially valid even through the subsequent data updates; the four-lepton lines refer to exotic channels discussed later). Right: Values of $B_e\cdot \sigma v$ (right axis) and of the boost factor $B_e$ (left axis, for $\sigma v = 3~10^{-26}{\rm cm}^3/{\rm sec}$) needed to fit the data (figure from~\cite{Cirelli:2008pk}).
\label{fig:fits}}
\vspace{0.2cm}
\end{figure}
Fig.~\ref{fig:fits} illustrates these points in a systematic way. On the left, it shows how the DM DM $\to \tau^+ \tau^-$ has the best $\chi^2$; other leptonic channels (e.g. $\mu^+\mu^-$) can give acceptable fits, but all other annihilations into quarks, vector and Higgs bosons are significantly disfavored. 
The value of the required annihilation cross section as a function of the DM mass is illustrated in fig.~\ref{fig:fits} (right).
The actual best fit case is illustrated in fig.~\ref{fig:chargedfit}: it consists of a candidate with a mass of 3 TeV and annihilating into $\tau^+ \tau^-$, a channel which produces smooth leptonic spectra, with a cross section of $2 \cdot 10^{-22}\, {\rm cm}^3/{\rm sec}$. 

\medskip

The appearence of a small but visible flux of antiprotons from a DM DM $\to \tau^+\tau^-$ annihilation mode in fig.~\ref{fig:chargedfit} (center) may be at first sight surprising. It is due to the fact that these fluxes are computed including {\bf electroweak corrections}, i.e. the radiation from the initial $\tau^\pm$ of EW gauge bosons ($W^\pm$, $Z$) which then decay into many other SM particles, including quarks that hadronize into antiprotons. More generally, the importance of such corrections has been appreciated only relatively recently, in a string of papers with varying scopes and levels of accuracy~\cite{EWcorr1}. Without entering in the details, it is enough for my purposes to remind that (i) the corrections are particularly relevant for large DM masses (above a TeV); (ii) they can alter significantly the ID fluxes, both in their spectral shape and in their amplitude, affecting especially the low energies portion~\cite{EWcorr2}, and (iii) in some cases they can also largely modify the annihilation cross section itself, since they can lift the helicity suppression into light fermions~\cite{EWcorr3}.

\medskip

Before moving on, one general remark is in order. As discussed, the excesses in $e^+$ and $e^-$ are real. If they are due to DM, great! If instead they are due to something else, then this `something else' represents a formidable background for any signal from DM in these species, which can be effectively forgotten. On the other hand, the exquisitely precise {\bf antiproton} data can be used to impose {\bf constraints} and, in perspective, one can hope to see some DM signal in them. This is what has been done in~\cite{Cirelli:2013hv}, finding that the current constraints are among the most stringent ones, although they are plagued by large astrophysical uncertainties. 

\medskip

This concludes my overview of the phenomenological interpretation of charged CR data. A discussion of how natural or preposterous the properties in page~\pageref{properties} are and of what it takes to realize them is postponed to Sec.~\ref{theory}. Here we proceed along the lines of a phenomenological model-independent approach.

\subsection{Photons}
\label{photons}

Given these tantalizing but surprising hints of Dark Matter annihilations in charged CRs, it is now crucial to consider the associated signals in the photon fluxes that necessarily accompany them. In general, these photon fluxes can be produced by DM in different ways, among which:
\begin{itemize}
\item[I)] {\bf `Prompt' gamma-rays}: produced directly by DM annihilations themselves. In turn, however, these gamma-rays can originate from different stages of the annihilation process:
\begin{itemize}
\item[Ia)] From the bremsstrahlung of charged particles and the fragmentation of hadrons, e.g. $\pi^0$, in the final states of the annihilations. These processes generically give origin to a {\it continuum} of $\gamma$-rays which peaks at energies somewhat smaller than the DM mass $m_{\mbox{\tiny DM}}$, i.e.\ typically in the $\gamma$-ray energy range of tens of GeV to multi-TeV. The spectra can be computed in a model independent way (see e.g.~\cite{Cirelli:2010xx}), since all one needs to know is the pair of primary SM particles. \label{promptcontinuum}
\item[Ib)] From the {\it bremsstrahlung} from one of the {\it internal} particles in the annihilation diagram~\cite{Bringmann:2007nk}. This typically gives rise to a sharp feature that peaks at an energy corresponding to the DM mass. The process is in general subdominant with respect to the continuum, but it can be particularly important in cases in which the continuum itself is suppressed by some mechanism, e.g. helicity arguments, which are lifted by the internal radiation. The spectrum from this contribution cannot be computed without knowing the details of the annihilation model. \label{promptIB}
\item[Ic)] From an annihilation directly into a pair of gamma-rays, which gives rise to a {\it line} spectrum at the energy corresponding to the mass of the DM. Since DM is neutral, this annihilation has to proceed via some intermediation (typically a loop of charged particles) and it is therefore suppressed by (typically) 2 to 4 orders of magnitude. \label{promptline}
\end{itemize}
In any case, since these $\gamma$-rays originate directly from the annihilations themselves, their spatial distribution follows closely the distribution of DM.

\item[II)] {\bf ICS gamma-rays}: produced by the Inverse Compton Scattering (ICS) of the energetic electrons and positrons, created in the DM annihilation, onto the low energy photons of the CMB, the galactic star-light and the infrared-light, which are thus upscattered in energy. Typically, they cover a wider range of energies than prompt gamma rays, from energies of a fraction of the DM mass to almost up to the DM mass itself. Their spatial distribution traces the distribution of $e^\pm$, which originate from DM but then diffuse out in the whole galactic halo (as seen above).

\item[III)] {\bf Synchrotron emission}: consisting in the radiation emitted in the magnetic field of the Galaxy by the $e^\pm$ produced by DM annihilations. For an intensity of the magnetic field of $\mathcal{O}$($\mu$Gauss), like in the case of the Milky Way halo, and for $e^\pm$ of GeV-TeV energies, the synchrotron emission falls in the MHz-GHz range, i.e. in the radio band. For large magnetic fields and large DM masses it can reach up to EHz, i.e. the X-ray band~\cite{Bergstrom:2006ny,Regis:2008ij}. Their region of origin is necessarily concentrated where the magnetic field is highest; in particular the Galactic Center is the usual target of choice. However, it has been recently suggested that the galactic halo at large, or even the extragalactic ones, can be interesting sources~\cite{Fornengo:2011iq}.
\end{itemize}

\bigskip

Individuating the {\bf best targets} to search for these annihilation signals is one of the main games in the field. Not very surprisingly, the preferred targets have to be {\bf (i)} regions with high DM densities {\em and/or}\  {\bf (ii)} regions where the astrophysical `background' is reduced and therefore the signal/noise ratio is favorable. The distinction between (i) and (ii) is of course not clear-cut, and of course there are specific cases in which other environmental reasons make a region more suitable than another (such as in the case of synchrotron radiation which needs a region with a strong magnetic field). Moreover, new promising targets keep being individuated. However, for the sake of schematizing, one can list the following targets at which the experiments look:
\begin{itemize}
\item[$\circ$] The Milky Way Galactic Center (GC) $-$ {\bf (i)} 
\item[$\circ$] Small regions around or just outside the GC, such as the Galactic Ridge (GR, an area enclosed within galactic longitude $-0.8^\circ<\ell < 0.8^\circ$ and latitude $|b| < 0.3^\circ$), the Galactic Center Halo (GCH, an annulus of about $1^\circ$ around the GC, excluding the Galactic Plane) etc $-$ {\bf (i)} + {\bf (ii)}
\item[$\circ$] Wide regions of the Galactic Halo (GH) itself (such as observational windows centered at the GC and several tens of degrees wide in latitude and longitude, or the so called `intermediate-latitude strips' defined by $10^\circ < b < 20^\circ$, or the Galactic Poles at $b > 60^\circ$), from which a diffuse flux of gamma-rays is expected, including the one due to the ICS emission from the diffused population of $e^\pm$ from DM annihilations $-$ {\bf (ii)}
\item[$\circ$] Globular clusters (GloC), which are dense agglomerates of stars, embedded in the Milky Way galactic halo. They are a peculiar kind of target since they are not supposed to be DM dominated, quite the opposite, as they are rich of stars. The interest in them arises from two facts: that they may have formed inside a primordial DM subhalo and some of the DM may have remained trapped; that the density of baryonic matter may create by attraction a DM spike and thus enhance the annihilation flux$-$ {\bf (i)} 
\item[$\circ$] Subhalos of the galactic DM halo, the position of which, however, is of course not known a priori. In a similar class, Intermediate Mass Black Holes (IMBH) have recently attracted some attention~\cite{Bertone:2005xz}, because they could create around them spikes of DM $-$ {\bf (i)}. 
\item[$\circ$] Satellite galaxies of the Milky Way, often of the dwarf spheroidal (dSph) class, such as Sagittarius, Segue1, Draco and several others, which are star-deprived and believed to be DM dominated $-$ {\bf (i)} + {\bf (ii)}
\item[$\circ$] Large scale structures in the relatively nearby Universe, such as galaxy clusters (e.g. the Virgo, Coma, Fornax, Perseus clusters, and several others with catalog names that are less pleasant to write) $-$ {\bf (i)} + {\bf (ii)}
\item[$\circ$] The Universe at large, meaning looking at  the {\em isotropic} flux of (redshifted) $\gamma$-rays that come to us from DM annihilation in all halos and all along the recent history of the Universe. Often this flux is called `extragalactic' or `cosmological' $-$ {\bf (ii)}
\begin{itemize}
\item[$\cdot$] It has also been suggested that experiments look at the angular {\em anisotropies} of this (at first
order) isotropic flux of diffuse gamma rays, since they should exist if annihilations happen in granular cosmological halos~\cite{Ando:2005xg,Fornasa:2009qh,Hensley:2009gh,Cuoco:2010jb}. A similar program has been suggested at the Galactic scale~\cite{SiegalGaskins:2008ge,Fornasa:2009qh,Ando:2009fp,Cuoco:2010jb} (it may also be listed in the `subhalos' point above). First results are presented in~\cite{Ando:2013ff}.
\end{itemize}
\end{itemize}

\bigskip

Focussing on the range of energies above a GeV or so (i.e. proper gamma rays), the current {\bf main experiments} in the game are the FERMI satellite and the ground-based Imaging Atmospheric \v Cerenkov Telescopes (IACT).
\begin{itemize}
\item[$-$] The FERMI Large Area Telescope (LAT) has un unprecedented sensitivity to gamma rays across four orders of magnitude in energy (30 MeV to 300 GeV). The Collaboration has published online all the raw data, which have been used by a large number of indipendent authors. The Collaboration itself has performed DM searches looking over most of the sky for gamma-ray lines~\cite{FERMI_lines,FERMI_lines2}, looking at satellite galaxies~\cite{FERMI_satellites_2008,FERMI_satellites_2011}, at the isotropic diffuse background~\cite{FERMI_isotropic}, at clusters of galaxies~\cite{FERMI_clusters}, at diffuse gamma rays from wide regions of the galactic halo~\cite{FERMI_diffusehalo,FERMI_lines2} and searching for DM satellites~\cite{FERMI_DMsatellites}. Among the many analyses not directly addressing DM, but of interest for DM, it is worth mentioning the one for anisotropies in the isotropic flux~\cite{Cuoco:2011ng}, which shows a positive detection of angular power but consistent with an astrophysical source, e.g. blazars.
\item[$-$] The HESS telescope is mainly sensitive to gamma rays in a range from tens of GeV to tens of TeV, nicely completing the FERMI range towards larger energies. The Collaboration has performed a large number of studies which are relevant for Dark Matter searches. There are observations towards the Galactic Center~\cite{HESS_GC}, the Galactic Ridge~\cite{HESS_GR}, the Galactic Center Halo~\cite{HESS_GCH}, a couple of globular clusters~\cite{HESS_globularclusters}, a number of dwarf galaxies such as  Sagittarius~\cite{HESS_Sgr_dwarf}, Carina and Sculptor~\cite{HESS_Carina_Sculptor_dwarf} and Canis Major (in the case of the latter, assuming it indeed is a dwarf galaxy)~\cite{HESS_CanisMajor} and the Fornax galaxy cluster~\cite{HESS_Fornax}. HESS has also looked at possible signals from spikes of DM accumulated around Intermediate Mass Black Holes~\cite{HESS_IMBH}.
\item[$-$] The VERITAS telescope has observed a few dwarf spheroidal galaxies~\cite{VERITAS_dwarfs} and the Coma galaxy cluster~\cite{VERITAS_Coma}. Its predecessor Whipple had also looked at a couple of dSphs, clusters and a GloC~\cite{WHIPPLE}. The MAGIC telescope  has observed a few dwarf galaxies too~\cite{MAGIC_dwarfs} and the Perseus galaxy cluster~\cite{MAGIC_Perseus}.
\end{itemize}

Besides these searches, many indipendent works have analysed varying combinations of datasets, considered different targets and studied different models. 
For example, in~\cite{Bertone:2008xr} we analysed the HESS measurements of the emission from the GC, the GR and Sagittarius dSph, as well as the radio data from the GC, while in~\cite{Cirelli:2009vg} and~\cite{Cirelli:2009dv} we  focussed on FERMI measurements of diffuse galactic gamma rays. So did~\cite{Meade:2009iu} and~\cite{Papucci:2009gd}. Similarly, the implications of preliminary FERMI data from the GC were considered in~\cite{Cholis:2009gv}. FERMI and MAGIC dwarf galaxies data were studied in~\cite{Essig:2009jx}. Ref.~\cite{Brun:2010ci} searched for DM galactic subhalos with HESS.  Searches for gamma ray lines in FERMI data have been addressed in~\cite{Vertongen:2011mu} and in~\cite{GeringerSameth:2012sr} (looking at dSphs). Similarly, the sharp features expected from Internal Bremsstrahlung have been searched for in FERMI data by~\cite{Bringmann:2012vr}. 
A multimessenger analysis which consider several probes was presented in~\cite{Pato:2009fn}.
Recently, FERMI data on dwarf galaxies have been independently studied by~\cite{GeringerSameth:2011iw} (see also~\cite{Cholis:2012}), while galaxy cluster observations by FERMI are used in~\cite{Huang:2011xr}, \cite{Ando:2012vu} and~\cite{Han:2012uw}. Ref.~\cite{Feng:2011ab} analyses FERMI data of the same couple of globular clusters as HESS. And this is only a very partial list.

\subsubsection{Photon constraints}
\label{photonconstraints}

\begin{figure}[p]
\begin{center}
\includegraphics[width=0.45\textwidth]{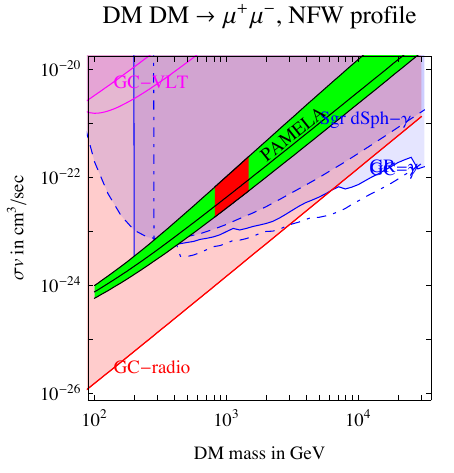} 
\includegraphics[width=0.41\textwidth]{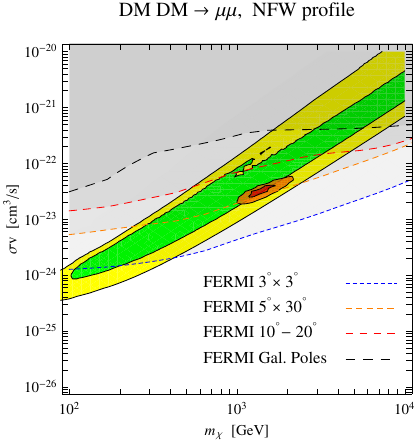} \\
\hspace{0.7cm}  \includegraphics[width=0.45\textwidth]{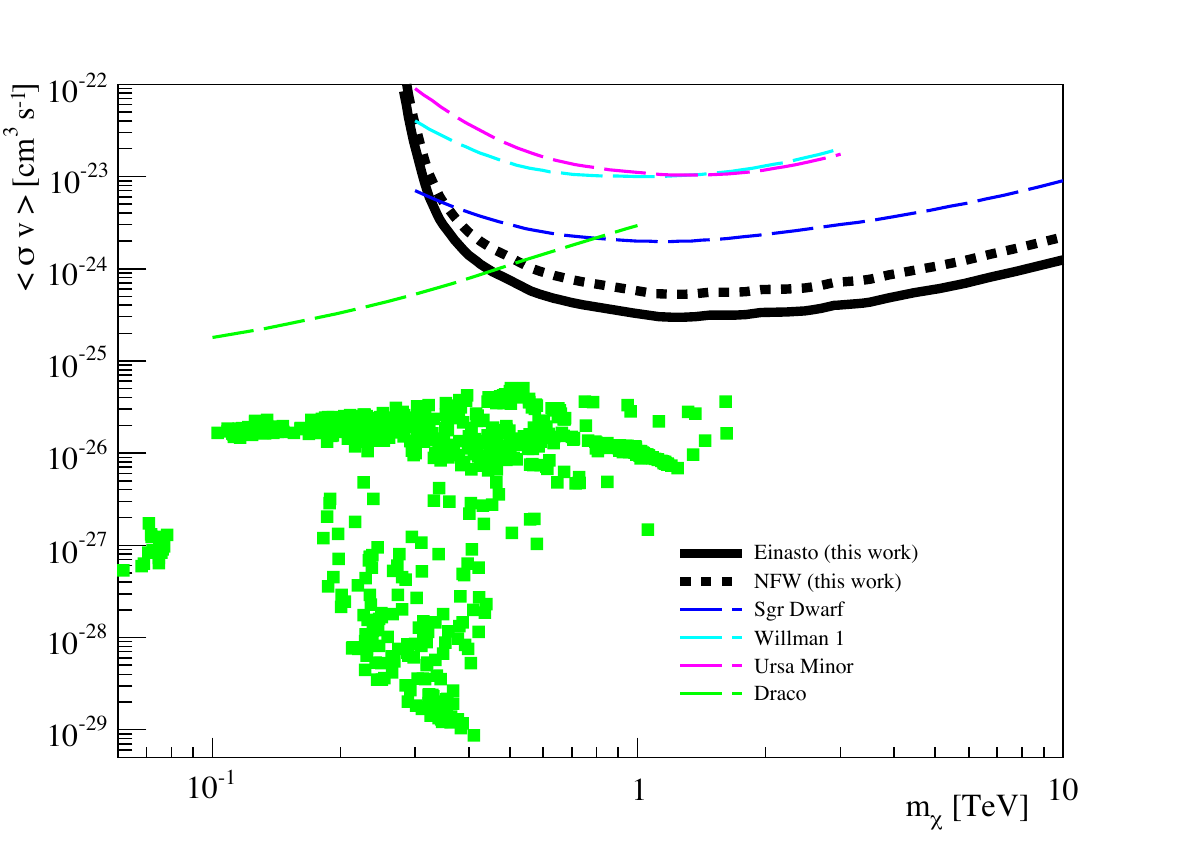} 
\includegraphics[width=0.45\textwidth]{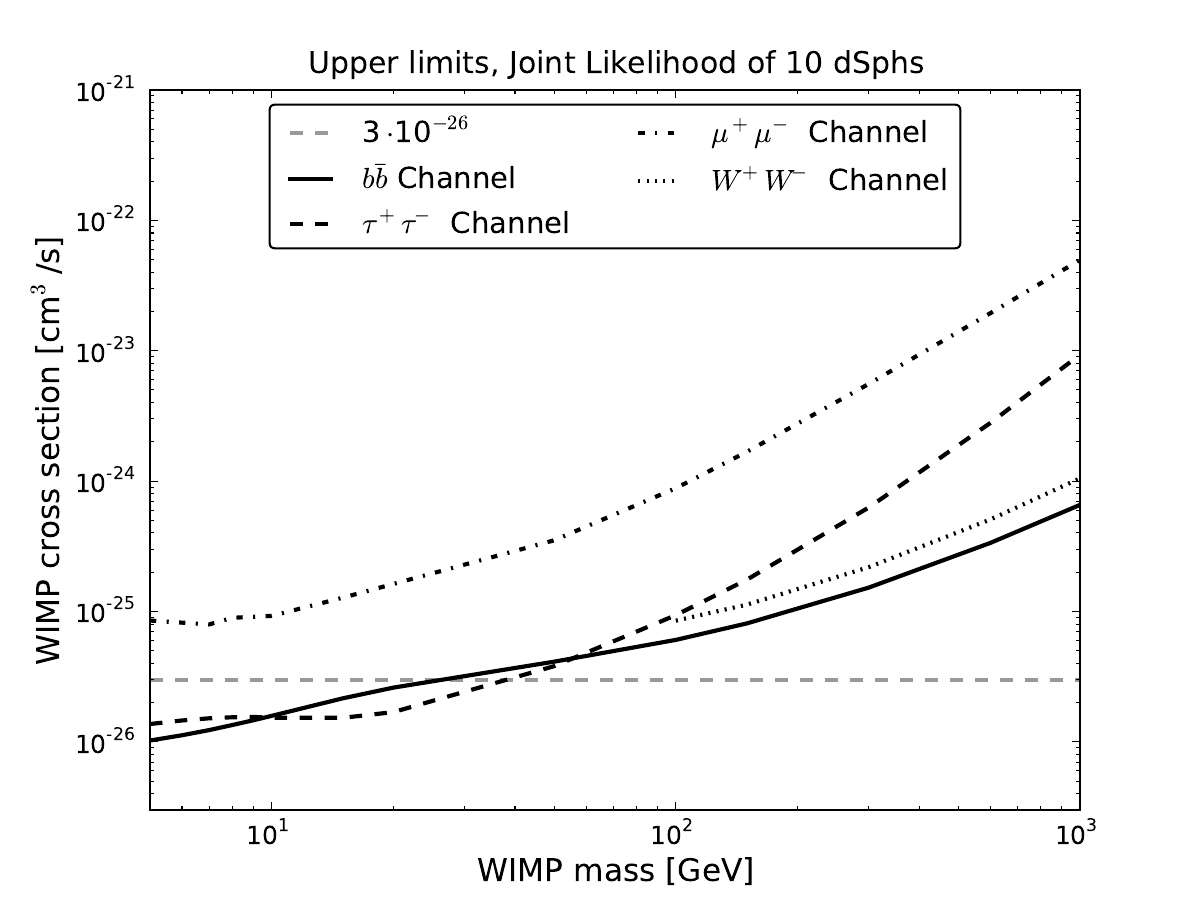} \\[0.1cm]
\includegraphics[width=0.65\textwidth]{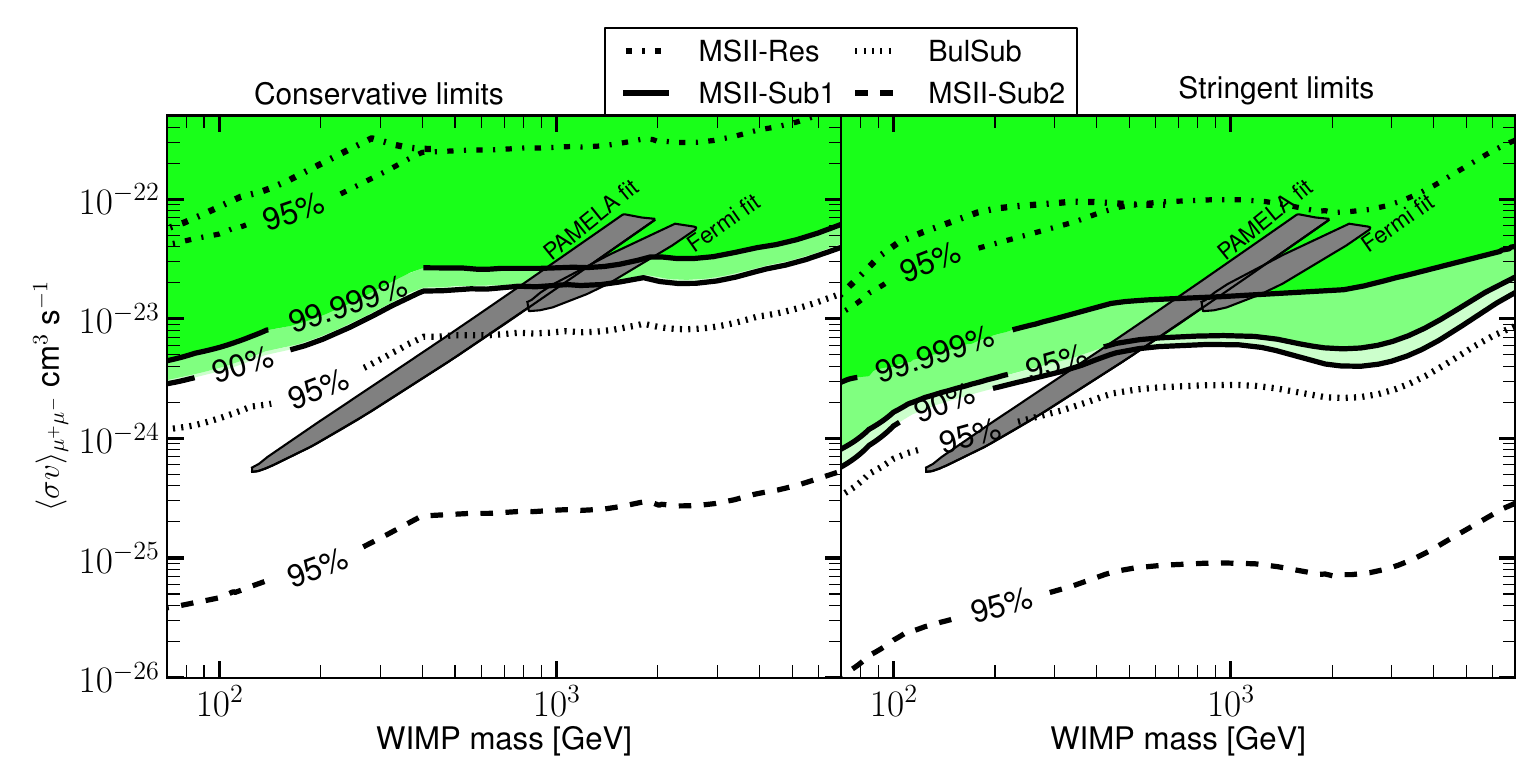}
\includegraphics[width=0.33\textwidth]{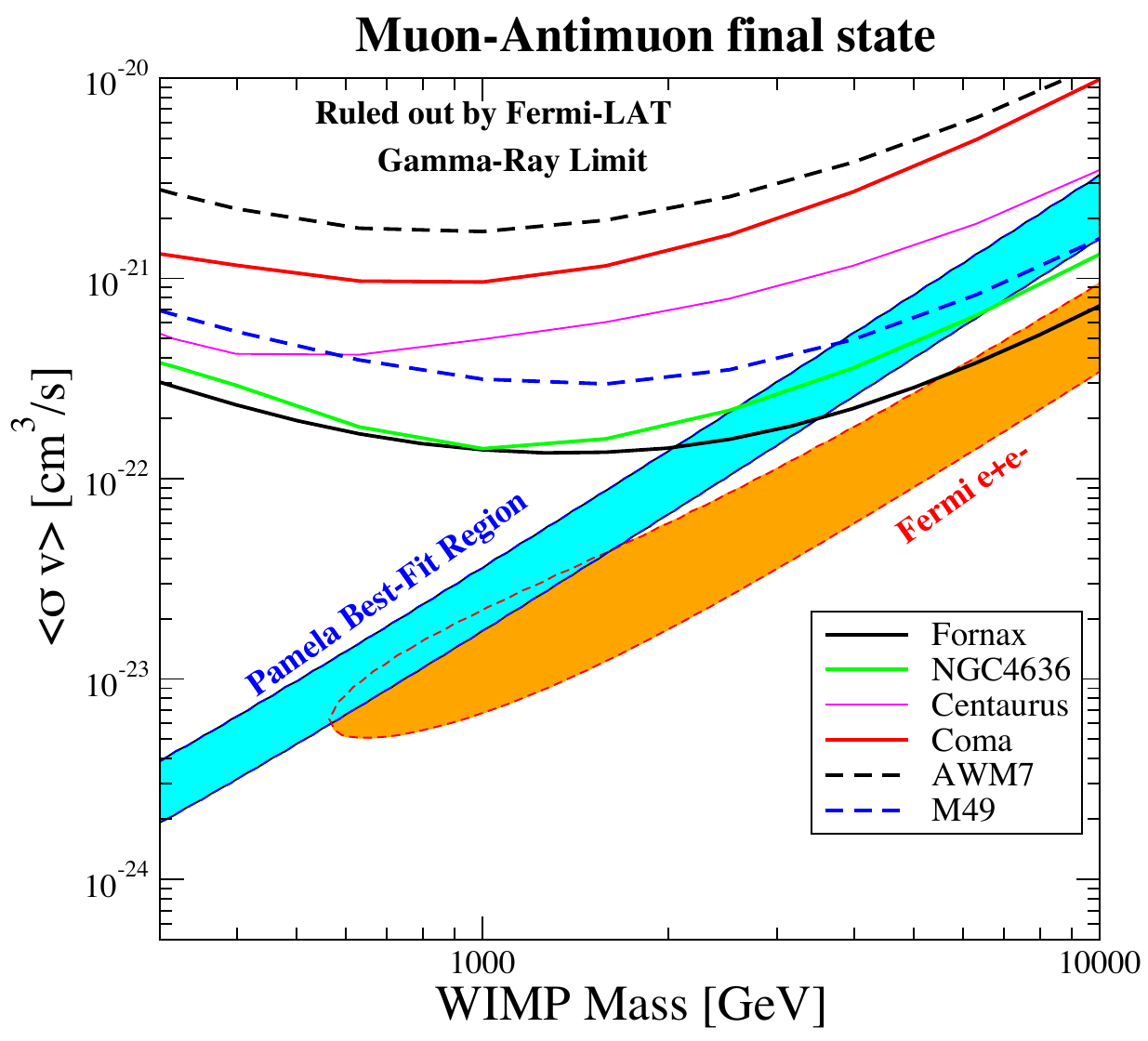}
\end{center}
\caption{A collection of recent bounds on DM annihilation from photon observations.
First row: Left: bounds from HESS observations of the Galactic Center (GC), Galactic Ridge (GR) and Sagittarius Dwarf (Sgr dSph) and from radio observations of the GC (figure from~\cite{Bertone:2008xr}). Right: bounds from FERMI observations of the Galactic Halo (from~\cite{Cirelli:2009dv}). 
Second row: Left: bounds from the HESS Coll. observation of the GCH (from~\cite{HESS_GCH}). Right: bounds by the FERMI Coll. from the observation of satellite galaxies (from~\cite{FERMI_satellites_2011}). 
Third row: Left: bounds from FERMI's measurement of the isotropic diffuse $\gamma$-ray background (from~\cite{FERMI_isotropic}). Right: bounds from FERMI's observation of galaxy clusters (from~\cite{FERMI_clusters}). For the sake of comparison, all plots refer to DM DM $\to \mu^+\mu^-$ annihilations (except for the HESS Coll.'s GC one which refers to `quark-antiquark' annihilation), but, most often, the corresponding references provide bounds for other annihilation channels. 
\label{fig:gamma}}
\vspace{-0.5cm}
\end{figure}

Without entering in the details of each single analysis, the overall common conclusions of almost all the studies cited above is: {\bf no ``anomalous'' signals} are individuated, therefore {\bf upper bounds on DM annihilation cross section} can be derived (see however Sec.~\ref{gammaclaims}). Fig.~\ref{fig:gamma} collects a few plots of such bounds.
\begin{figure}[t]
\begin{center}
\includegraphics[width=0.32\columnwidth]{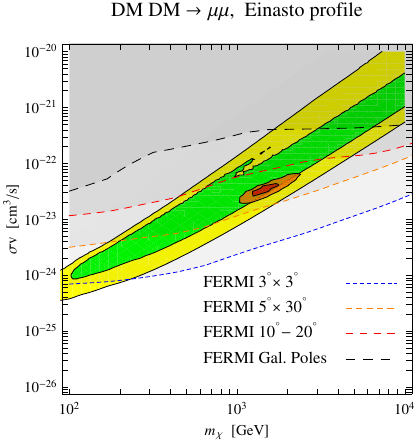} \ \
\includegraphics[width=0.32\columnwidth]{gamma_mybounds_ICS_NFWmu} \ \ 
\includegraphics[width=0.32\columnwidth]{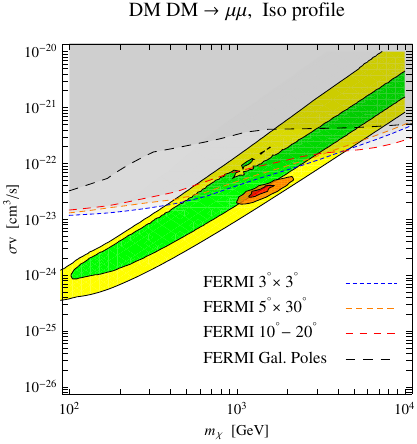}
\caption{Illustration of a typical dependence of GH $\gamma$-ray bounds on the choice of DM profile. Having fixed the particle physics model, we change the DM profile of the MW from Einasto (left) to NFW (center) to Isothermal (right). For the most shallow profiles the gamma ray bounds from the inner regions are lifted and the PAMELA+FERMI+HESS fit area (orange) is reallowed. Figure from~\cite{Cirelli:2009dv}.}
\label{fig:gammachangeprofile}
\end{center}
\end{figure} 

\noindent Now, therefore, for the flow of our discussion two main questions need to be addressed: 
\begin{itemize}
\item[$\triangleleft$] 1. Are these constraints enough to rule out the DM interpretation of charged CRs excesses, i.e. a few-TeV, leptophilic DM with very large annihilation cross section?
\item[$\triangleleft$] 2. Are these constraints strong enough to test the standard thermal DM production mechanism?
\end{itemize}
The answers are not as clear-cut as one would hope:
\begin{itemize}
\item[$\triangleright$] 1. {\em Yes, unless...} Yes, several indipendent constraints rule out the portion of the $m_{\mbox{\tiny DM}}-\sigma v$ parameter space preferred by the charged CR fits, identified for instance in fig.~\ref{fig:gamma} by red, orange or gray blobs. Yet, there remain specific assumptions for which a marginal compatibility can be found. Most notably concerning the choice of DM profile in the Galaxy: if it is cored, something which is however disfavored by numerical simulations, the flux of photons from the inner regions of the Galaxy is expected to be lower and the constraints are somewhat lifted. This simple point is illustrated in fig.~\ref{fig:gammachangeprofile} in the case of constraints from the diffuse gamma rays from the GH. 
\item[$\triangleright$] 2. {\em No, except...} No, the current bounds lie from a factor of a few to several orders of magnitude higher than the `natural' annihilation cross section $\sigma v = 3 \cdot 10^{-26}$ cm$^3$/s (see Sec.~\ref{wimp}). One exception is the recent constraint from satellite dwarf galaxies with FERMI data~\cite{GeringerSameth:2011iw}\cite{FERMI_satellites_2011}: for some annihilation channels, thermal DM is excluded for masses below 30 GeV or so (see plot in fig.~\ref{fig:gamma}). Another exception is constituted by the bounds from globular clusters derived in~\cite{Feng:2011ab}, which are even stronger if taken at face value, but which crucially depend on the poorly known DM distribution in the GloC and its history. Another recent exception is~\cite{Han:2012uw}, which rules out thermal DM up to $\sim$100 GeV using observations of the Virgo cluster. 
\end{itemize}
Nevertheless, the bottom line message is clear: in most $\gamma$-ray searches, no signal of DM is seen and a tension with charged CRs (which goes from fatal to somewhat uncomfortable) is present.

\subsubsection{A scent of DM... in photons?}
\label{gammaclaims}

Actually, sporadically, claims of evidence of DM in $\gamma$-rays data from the FERMI satellite have been made. 
This is currently a hot topic, so let me start with a {\it Wikipedia}-like caveat:
\vspace{-0.6cm}
\begin{figure}[!h]
\begin{minipage}{0.1\textwidth}
\begin{flushleft}
\includegraphics[width=\textwidth]{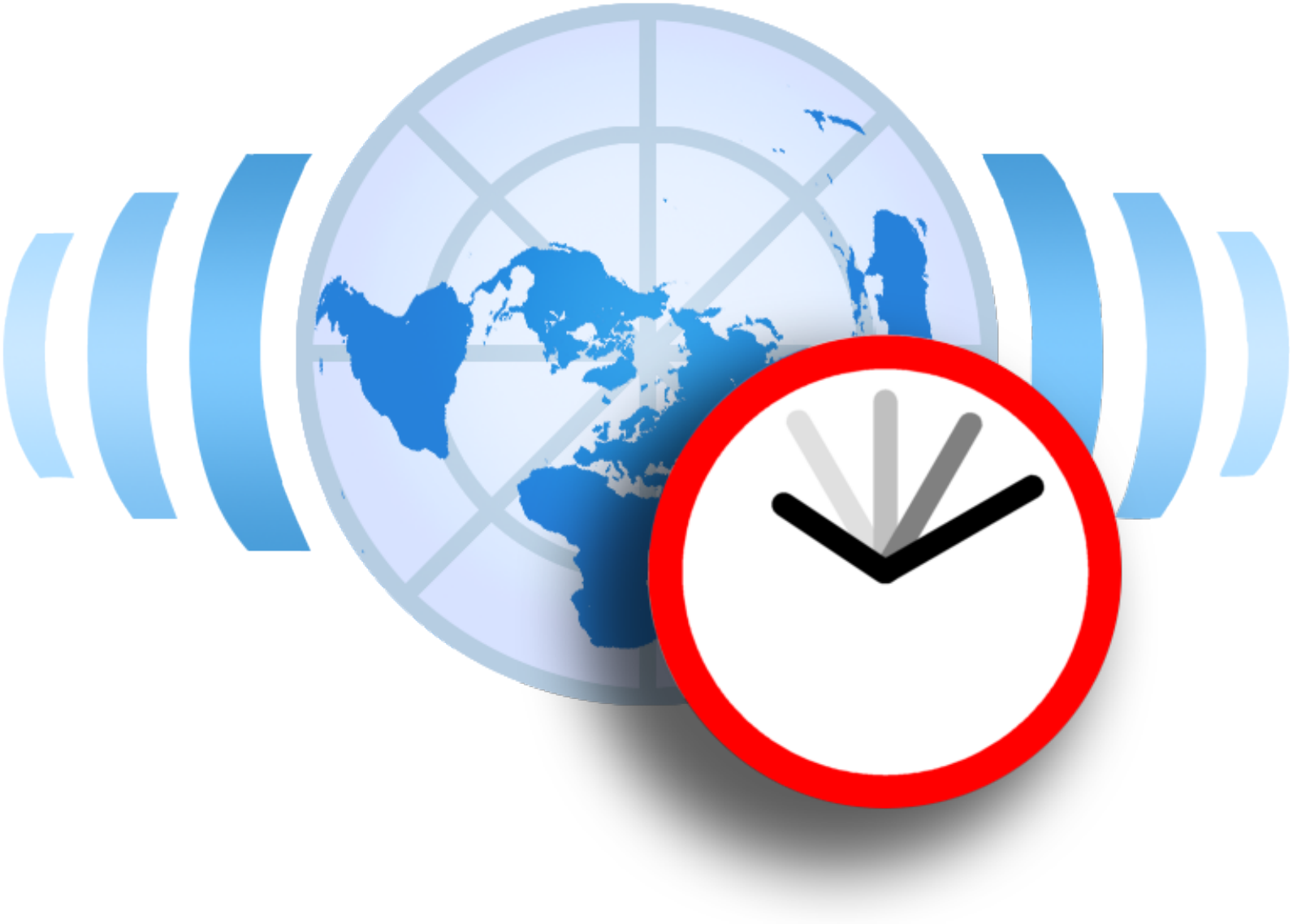} 
\end{flushleft}
\end{minipage}
\quad
\begin{minipage}{0.8\textwidth}
\begin{flushleft}
{\bf This section documents a current event.}\\ Information may change rapidly as the event progresses. {\footnotesize \it (August 2012)}
\end{flushleft}
\end{minipage}
\end{figure}

\vspace{-0.3cm}
Evidences for DM have been reported at the GC in~\cite{Goodenough:2009gk,Hooper:2010mq,Hooper:2011ti},\cite{Abazajian:2012pn}, in the isotropic flux~\cite{Hutsi:2010ai}, in the Virgo cluster~\cite{Han:2012au}, possibly in the MW halo~\cite{Lin:2010fba}... Most notably and most recently an evidence for one (or two) $\gamma$-ray line(s) around 130 GeV has been reported, as discussed below. 

Setting aside for a moment this latter claim, it is fair to say that none of the other ones, so far, has gathered enough consensus to be considered more than a tentative hint. The crucial points of criticism often have to do with the assumptions made for the astrophysical background (which can, at this stage of significance, mimic a DM signal)
%, see e.g.~\cite{Abazajian:2010zy,Boyarsky:2010dr} addressing the claim in~\cite{Hooper:2010mq} and the rebuttal in~\cite{Hooper:2011ti,Hooper:2012ft} 
or the lack of independent confirmations of the same signal in other channels or in other locations. 

In this respect there are two strings of papers that illustrate well the difficulty of the analyses and of the debate. (i) The claim in~\cite{Hooper:2010mq} in favor of light ($\mathcal{O}$(10) GeV) DM annihilations at the GC has been questioned in~\cite{Abazajian:2010zy,Boyarsky:2010dr}, which essentially affirm that known astrophysical sources (e.g. millisecond pulsars, MSP) can explain it away. But it has then been re-claimed in~\cite{Hooper:2011ti,Hooper:2012ft}, saying that those sources are too few, too point-like, too dim or too soft to produce the emission. The latest installment consists of~\cite{Abazajian:2012pn}, which reaffirms the MSP hypothesis while entertaining the DM one as well. 
(ii) The claim in~\cite{Han:2012au} of DM annihilations in the Virgo cluster has been refuted first in~\cite{MaciasRamirez:2012mk} and then by an extended set of authors of the original paper in~\cite{Han:2012uw}. In particular, the latter found that, removing previously-unaccounted-for astrophysical point sources, the evidence in~\cite{Han:2012au} evaporates and actually one can impose stringent bounds (possibly the most stringent ones) using the very same Virgo cluster observation.  

\medskip

Let's now come to `130 GeV line' claim (for a more thorough review see~\cite{reviewBringmannWeniger}). Originally spotted by~\cite{Bringmann:2012vr} and, above all, by~\cite{Weniger:2012tx} in the publicly available FERMI data from an extended region including the GC, it has later found support in other analyses~\cite{Tempel:2012ey,Su:2012ft,Hektor:2012kc,Su:2012zg}, with varying degrees of accuracy and claimed significance. \cite{Tempel:2012ey,Su:2012zg} have seen it in what could possibly be DM subhaloes of the MW, and there might be two lines, at 111 GeV and 129 GeV~\cite{Rajaraman:2012db,Su:2012ft}. \cite{Hektor:2012kc} has seen it in galaxy clusters too. For a response,~\cite{Boyarsky:2012ca,Mirabal:2012za,Hektor:2012jc} challenged the analyses in a number of ways, suggesting that the line(s) could be due to unidentified instrumental, statistical or astrophysical origin.~\footnote{By the way, can astrophysics produce a $\gamma$-ray line (something which has been advertised for decades as a `smoking gun' evidence for DM)? Yes, of course, it sufficed to ask and pulsars (always pulsars...) took on themselves the task of producing lines at 100 GeV or somewhat above, via cold electron winds impinging on environmental gamma radiation~\cite{Aharonian:2012cs}. It is challenging, though, to see how this process could give rise to a spatially extended emission.} If however the line(s) is (are) from DM, it is plausible to expect an associated $\gamma$-ray continuum (and possibly a flux of other CRs, e.g. anti-protons) at lower energies, originating from the annihilations of DM into other SM particles; if the cross section into $\gamma\gamma$ is normalized to the one required by FERMI data ($\mathcal{O}(10^{-27})\, {\rm cm}^3/{\rm s}$), the inferred flux in the continuum poses problems in a variety of cases~\cite{Buchmuller:2012rc,Cohen:2012me,Cholis:2012fb,Huang:2012yf,Hooper:2012qc}. However, the details are model dependent: there is no airtight exclusion and actually working exceptions can be built~\cite{Tulin:2012uq}.

\medskip

In any case, it is clear that this is one of the most interesting areas of development and no claim should be dismissed without deep scrutiny. It might well be that one of these claims will prove to be the harbinger of a full-fledged discovery!

In general, some of our best hopes for clarifying the situation lie in the FERMI satellite itself, of which the prelaunch predicted sensitivities~\cite{Baltz:2008wd} let us believe that a DM with thermal cross section will be probed at 3$\sigma$ up to a mass of several tens or even hundreds of GeV, depending on the annihilation channel and the chosen target.
Another set of upcoming experiments will further improve the sensitivity on $\gamma$-ray lines, e.g. HESS-II which will soon be operational, CTA which should start in a few years and Gamma-400 which is being advocated for. In the merit of the 130 GeV line signal, they will provide precious information~\cite{Bergstrom:2012vd}.

\subsection{Neutrinos}
\label{neutrinos}

Neutrinos are of course produced in DM annihilations together with all the other particles discussed above. Similarly to $\gamma$-rays, neutrinos have the advantage of proceeding straight and essentially unabsorbed through the Galaxy. Even more, they can cross long lengths of dense matter with little interaction. Contrary to $\gamma$-rays, however, the {\bf detection principle} of neutrinos is more difficult and it introduces limitations in the choice of targets. Neutrinos are observed at huge \v Cerenkov detectors located underground (or under-ice or under-water) via the showers of secondary particles that they produce when interacting in the material inside the instrumented volume or in its immediate surroundings. The charged particles, in particularly muons, emit \v Cerenkov light when traversing the experiment and thus their energy and direction (which are connected to those of the parent neutrino) can be measured. The main background for this search consists in the large flux of cosmic muons coming from the atmosphere above the detector. The experiments, therefore, have to select only upgoing tracks, i.e. due to neutrinos that have crossed the entire Earth and interacted inside or just below the instrumented volume. 

\begin{figure}[p]
\begin{center}
\includegraphics[width=0.52\textwidth]{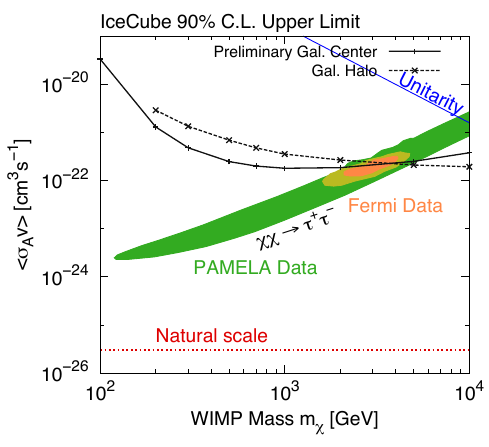}
\includegraphics[width=0.47\textwidth]{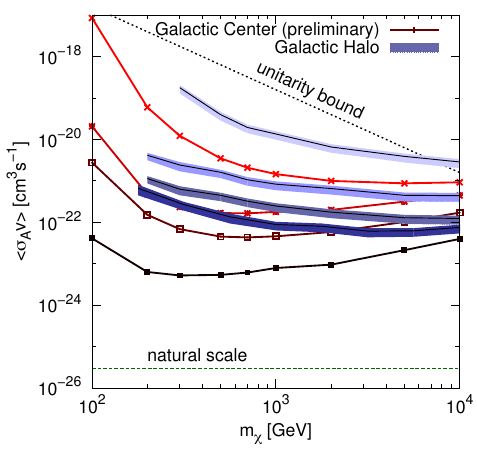}\\
\includegraphics[width=0.49\textwidth]{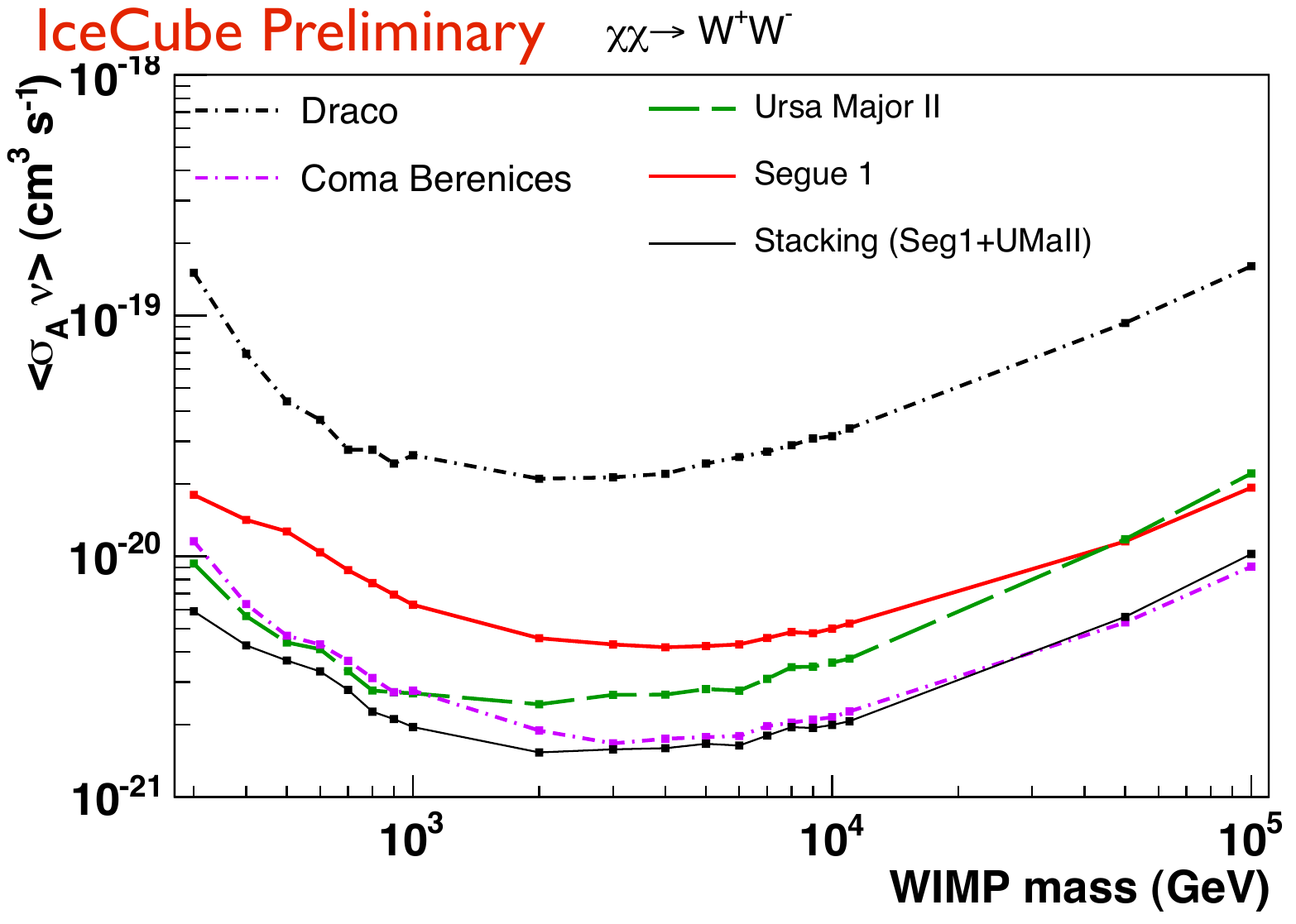}
\includegraphics[width=0.49\textwidth]{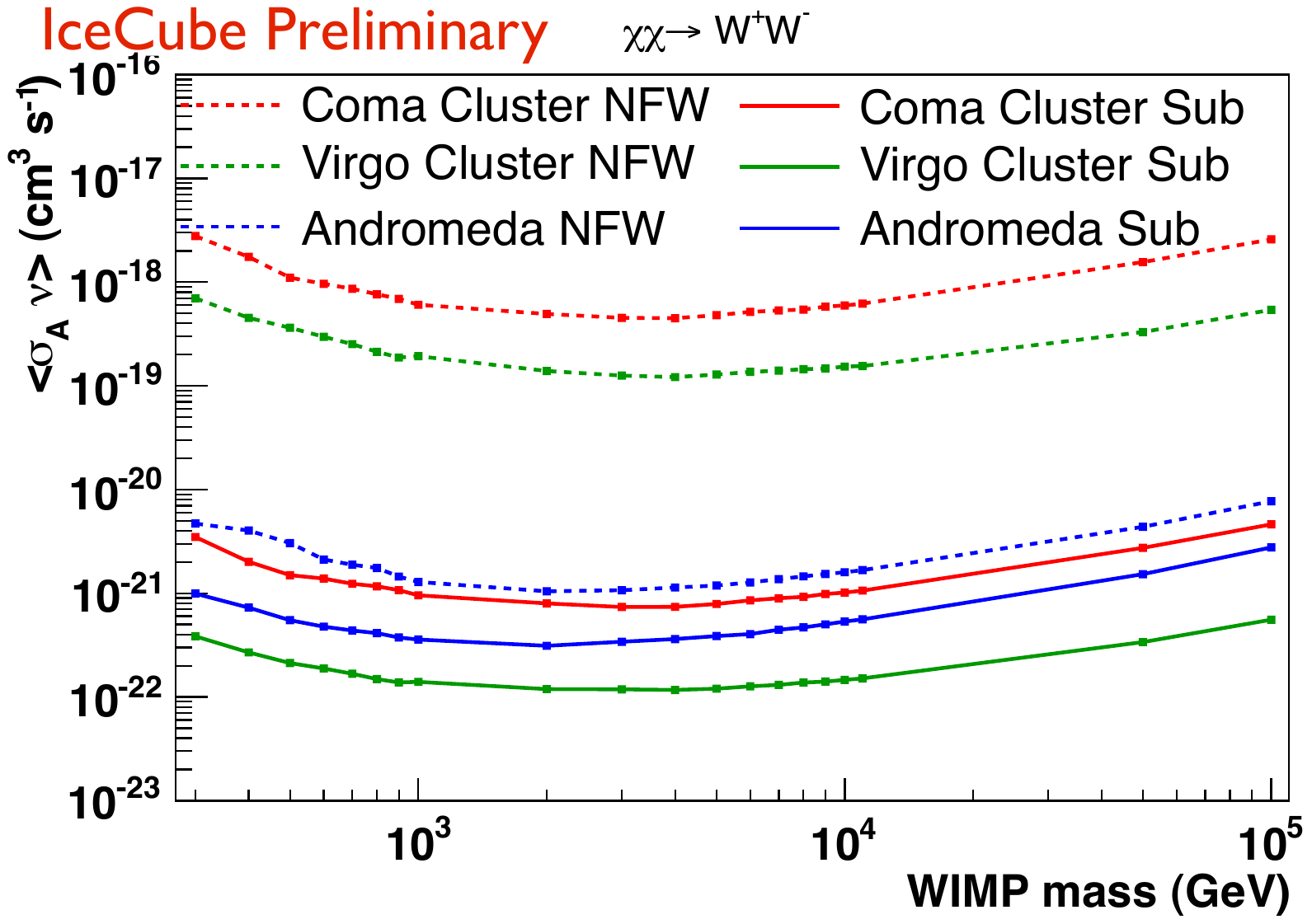}\\[-0.5cm]
\includegraphics[width=0.54\textwidth]{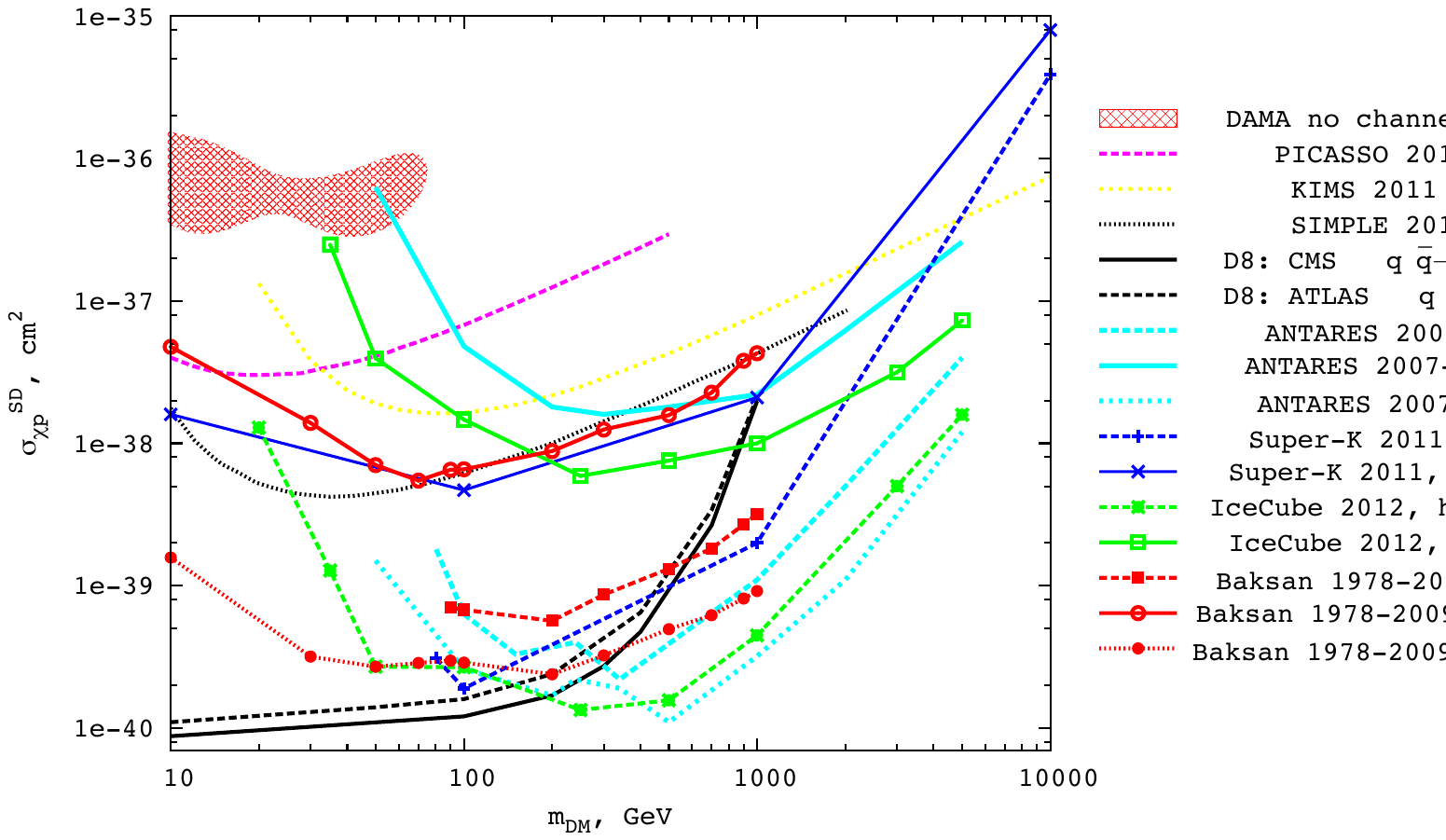}\hspace{0.3cm}
 \raisebox{0.5cm}{\includegraphics[width=0.42\textwidth]{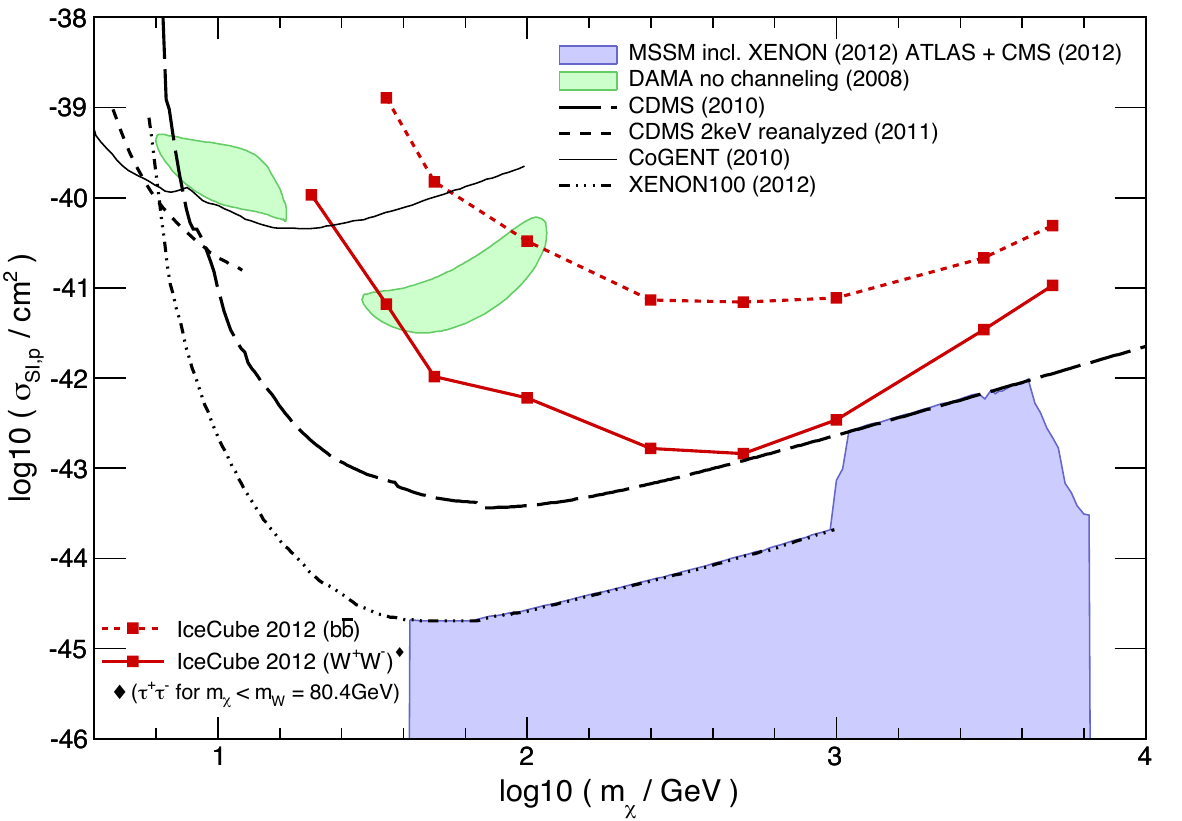}}
\end{center}
\vspace{-0.5cm}
\caption{A collection of recent bounds on DM from neutrino observations. 
First row: Left: constraints on the DM annihilation cross section from ICECUBE's observation of the Galactic Halo, comparing with the fit regions of charged CRs (figure from~\cite{IceCube:2011ae} or~\cite{Abbasi:2011eq}). Right: a compilation of current constraints from ICECUBE, from the GC and the galactic halo (figure from~\cite{IceCube:2011ae} or~\cite{delosheros}).
Second row: Left: ICECUBE constraints from dwarf galaxies (figure from~\cite{RottiDM2012}. Right: ICECUBE constraints from galaxy clusters (figure from~\cite{RottiDM2012}.
Third row: Left: bounds on the DM spin-dependent scattering cross section on nuclei from SuperKamiokande's, ICECUBE's and Baksan's searches for high-energy neutrinos from the Sun (from~\cite{Boliev:2013ai}). Right: the same for spin-independent scattering (from~\cite{icecube:2012ef}).  
\label{fig:neutrinos}}
\vspace{-0.3cm}
\end{figure}

Schematically, experiments look for neutrinos:
\begin{itemize}
\item[$\circ$] From the GC or the GH, in close similarity with $\gamma$-rays. Experiments located at the South Pole can not `see' the GC, which is essentially above horizon for them. The DeepCore extension of ICECUBE, however, circumvents this limitation by using the outer portion of the experiment as an active veto.
\item[$\circ$] From satellite galaxies or clusters of galaxies, again in similarity with $\gamma$-rays, although in this case the sensitivities are not competitive with gamma rays (at least unless one considers very large DM masses~\cite{Murase:2012rd}).
\item[$\circ$] From the center of the Sun (or even the Earth). The idea is that DM particles in the halo may become gravitationally captured by a massive body, lose energy via repeated scatterings with its nuclei and thus accumulate at its center. The annihilations occurring there give origin to fluxes of high energy neutrinos which, albeit suffering oscillations and interactions in the dense matter of the astrophysical body (see e.g.~\cite{DMnu,Blennow:2007tw}), can emerge. The detection of high-energy neutrinos from the Sun, on top of the much lower energy neutrino flux due to nuclear fusion processes, would constitute the proverbial smoking gun for DM, as there are no known astrophysical processes able to mimic it.
\end{itemize}

\subsubsection{Neutrino constraints}
\label{neutrinoconstraints}

The main neutrino telescopes, such as SuperKamiokande, ICECUBE and its predecessor Amanda, have looked for signals, without finding any. This therefore imposes once again bounds on the relevant DM properties. Fig.~\ref{fig:neutrinos} collects a few representative ones. 
%The ANTARES mediterranean telescope, another player in the game, has for the moment presented only projected sensitivities~\cite{ANTARES}.

The non-observation of high-energy neutrino fluxes from the GC imposes {\bf constraints on the DM annihilation cross section}. SuperKamiokande data have been originally analysed in~\cite{Desai:2004pq} and then more recently in~\cite{Meade:2009iu}. The ICECUBE collaboration also looked for neutrinos from the DM GH, from the GC~\cite{Abbasi:2011eq,delosheros} and recently from dwarf galaxies and clusters of galaxies~\cite{RottiDM2012}. Quantitatively, such constraints fall somewhat above the $\gamma$-ray bounds discussed in Sec.~\ref{photonconstraints}, and are therefore slightly less stringent. They have the advantage, however, of being less dependent on the DM particle mass: the reason is that while a larger DM mass implies lower DM density (e.g. in the GC) and therefore fainter fluxes and looser constraints, this is compensated by the fact that the higher energy neutrinos coming from such heavier DM annihilations also have a higher cross section for detection (in the material of neutrino telescopes) and thus the loss in the rate is partly compensated. Neutrino constraints become therefore somewhat competitive with $\gamma$-ray bounds at large DM masses.

The non-observation of high-energy neutrino fluxes from the center of the Sun, on the other hand, imposes {\bf constraints on the scattering cross section of DM particles with nuclei}, the same which are relevant for DM Direct Detection (DD). 
The ICECUBE and AMANDA collaborations present bounds in~\cite{IceCube:2011ae,IceCube:2011aj,RottiDM2012,icecube:2012ef}, SuperKamiokande presents results from 3109.6 days of data taking (!) in~\cite{Tanaka:2011uf}, updating~\cite{Desai:2004pq}, the ANTARES neutrino telescope presents an analysis of their early dataset in~\cite{Adrian-Martinez:2013tna} and finally the Baksan experiment presents results from 24.12 years of live-time (!!) in~\cite{Boliev:2013ai}.
Remarkably, the bounds are competitive with those from the dedicated DD experiments, such as Xenon100 or CDMS, both on the spin-dependent scattering cross section and on the spin-independent one. 
Ref.~\cite{Kappl:2011kz} has looked in particular at SuperKamiokande data from the Sun and their implication for light DM, of interest for explaining the events in DAMA, CoGeNT and possibly CRESST-II: for the spin-independent case, neutrino constraints exclude light DM if it annihilates into $\tau^+\tau^-$ or neutrino pairs, while for the spin-dependent case they exclude all channels.

\subsection{Antideuterons}
\label{antideuterons}

Antideuterons (the bound states of an antiproton and an antineutron) fall of course in the category of charged CRs, already discussed in~\ref{charged}. They get to get a subsection of their own because of two lame excuses: because they have not been detected yet, so there is no data-status to speak of, and because, on the other hand, they are believed to be quite promising as a tool for DM searches.

Antideuterons can be produced by Dark Matter via the {\bf coalescence} of an antiproton and an antineutron originating from an annihilation event, provided that the latter ones are produced with momenta that are spatially alligned and comparable in magnitude~\cite{Donato:1999gy}. This peculiar kinematics in the production mechanism implies two things. One, that the flux of $\bar d$ from DM is (unfortunately) predicted to be much lower than the one of other, more readily produced, charged cosmic rays. Two, that (fortunately) the flux peaks in an energy region, corresponding typically to a fraction of a GeV, where very little astrophysical background $\bar d$'s are present, since the latter ones are believed to originate in spallations of high energy cosmic ray protons on the interstellar gas at rest, a completely different kinematical situation. It is therefore sometimes said that the detection of even just one sub-GeV antideuteron in cosmic rays would be a smoking gun evidence for DM.

The GAPS~\cite{GAPS} and AMS-02~\cite{AMS02,AMS02foreseen} {\bf experiments} are going to look for this smoking gun. The GAPS experiment, in particular, will be a dedicated balloon or satellite mission which employs a novel technique: it plans to slow down the $\bar d$ nucleus, have it captured inside the detector to form an exotic atom and then annihilate emitting characteristic X-ray and pion radiation. Currently there is only an upper limit from the BESS experiment~\cite{BESSlimit}, at the level of about 2 orders of magnitude higher than the most optimistic predictions. 

\begin{figure}[t]
\begin{center}
\includegraphics[width=0.40\textwidth]{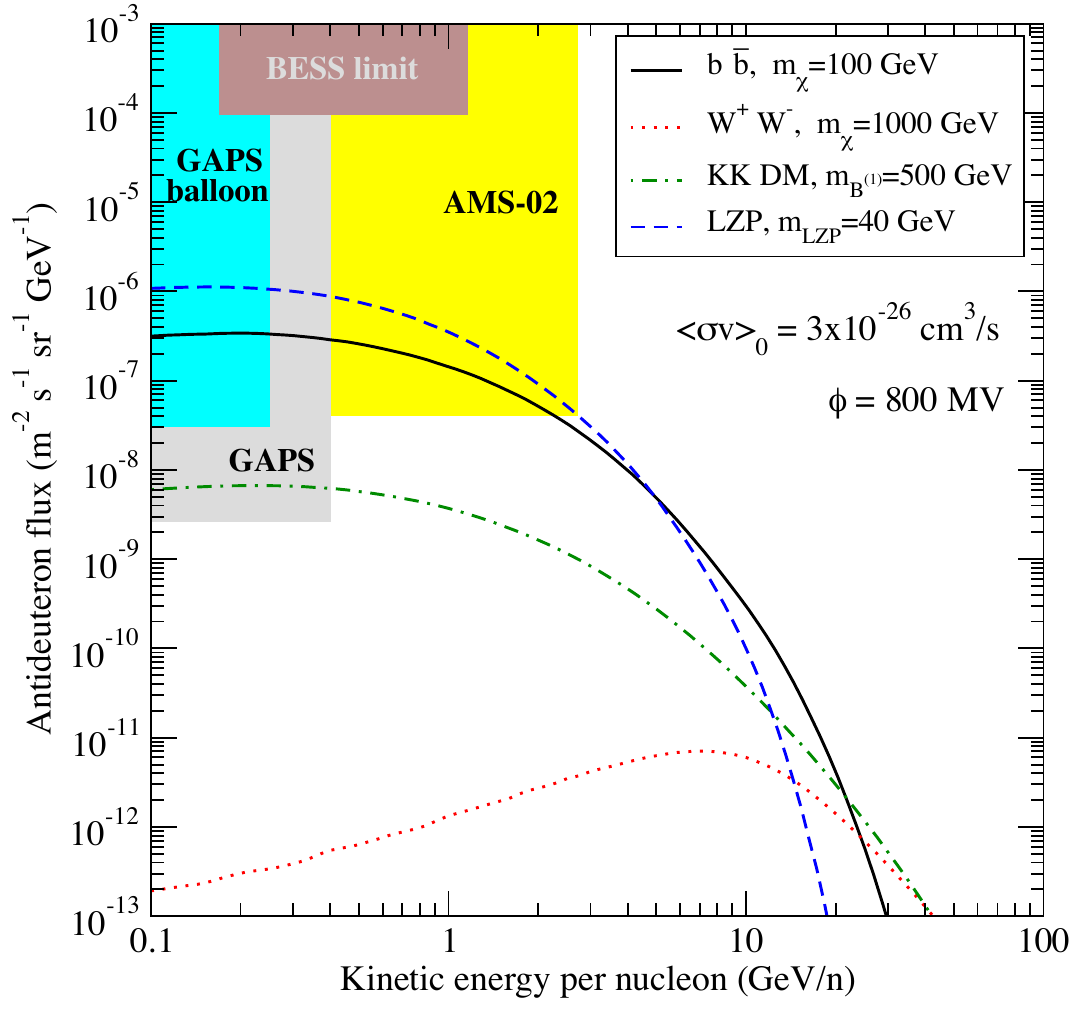} \hfill
\includegraphics[width=0.59\textwidth]{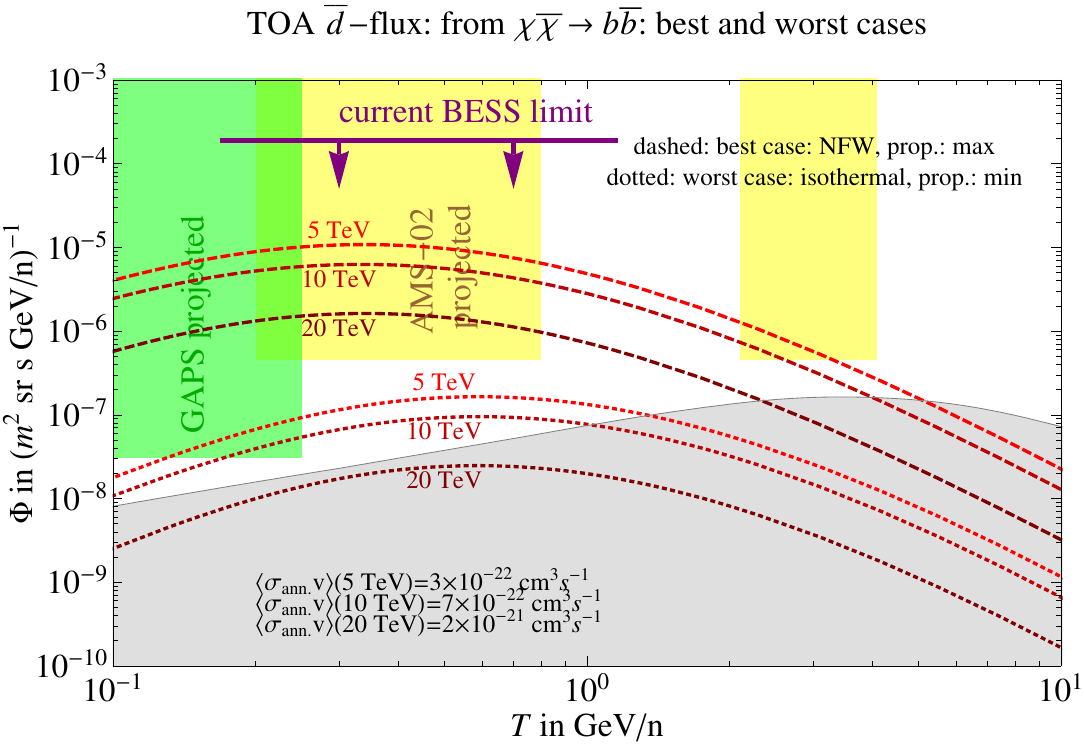}
\end{center}
\vspace{-5mm}
\caption{Some predictions for the flux of $\bar d$ from DM, compared with the current BESS limit and projected sensitivities of GAPS and AMS-02. Left: for typical DM models such as SuSy, KK DM etc (figure from~\cite{Baer:2005tw}). Right: for multi-TeV DM (from~\cite{Brauninger:2009pe}).  
\label{fig:antideuterons}}
\end{figure}

Recent analyses~\cite{Baer:2005tw,Donato:2008yx,Brauninger:2009pe,Kadastik:2009ts,Cui:2010ud} seem to suggest {\bf promising fluxes} in different scenarios, see e.g. fig.~\ref{fig:antideuterons}. Ref.~\cite{Kadastik:2009ts}, in particular, has pointed out that taking into account the fact that the $\bar p$ and the $\bar n$ are produced in a jet in the annihilation process boosts the coalesce rate and may give rise to a detectable signal also at much larger energies.

\subsection{Cosmology}
\label{cosmology}

\begin{figure}[t]
\begin{center}
\includegraphics[width=0.39\columnwidth]{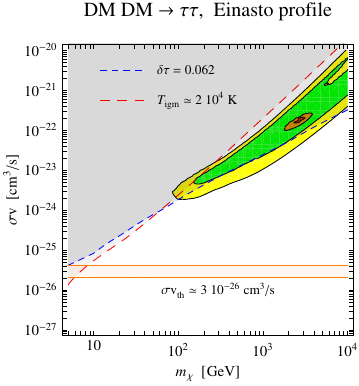} \hfill
\includegraphics[width=0.56\columnwidth]{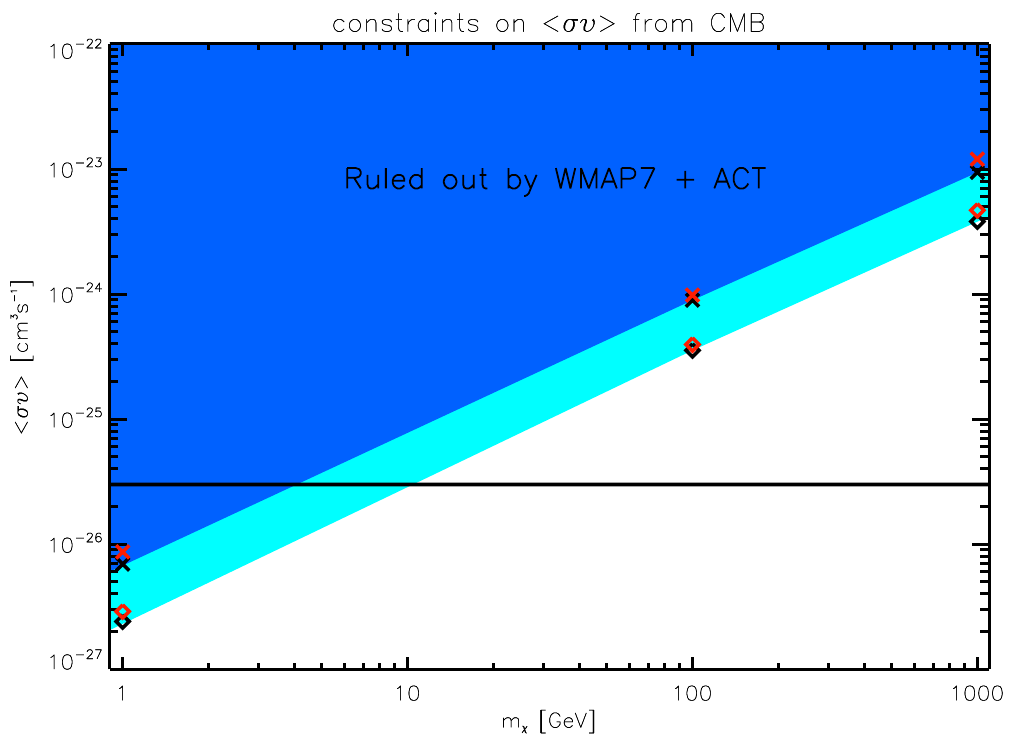}
\caption{A couple of representative bounds on DM annihilation from cosmology. Left: bounds from reionization, i.e. WMAP measurements of the optical depth of the Universe (blue dashed line, figure from~\cite{Cirelli:2009bb}). Right: most recent determination of the bounds from CMB (from~\cite{Galli:2011rz}).}
\label{fig:cosmology}
\end{center}
\end{figure}

Cosmology deserves a short paragraph of its own within a mini-review dedicated to Dark Matter Indirect Detection, because it can impose very relevant constraints on the annihilation cross section. This does not refer to the cosmological mechanism of production of DM (which can give at best a suggestion for a natural value of $\sigma v$, as discussed in the Introduction), but instead to later phases in the history of the Universe on which DM annihilations can leave an impact. 

The basic physics mechanism is that {\bf DM annihilations inject energy and energetic particles} in the primordial medium, and therefore affect its evolution. The constraints come essentially from two cosmological probes: Big Bang Nucleosynthesis (BBN) and the Cosmic Microwave Background (CMB). 

\begin{itemize}
\item[i)]
The injection of electromagnetic radiation or high-energy hadrons during or after {\bf BBN} can have several effects on the newly formed light nuclei, destroying some species and overproducing others~\cite{Jedamzik:2006xz,Jedamzik:2009uy}, therefore altering their final abundances and spoiling the agreement with observations. 
The bounds found in~\cite{Hisano:2009rc} vary in the range $10^{-23} \div 10^{-21} {\rm cm}^3$/sec (for a 1 TeV DM mass), depending on the DM annihilation channel and the light element which is used. They lie therefore quite above the value of the thermal cross section and they are also generically not strong enough to significantly constraint the DM interpretation of the charged CRs excesses (especially because leptophilic channels are constrained the least).

\item[ii)]
Concerning {\bf CMB}, the actual physical effect of energy injection around the recombination epoch is that it ionizes the gas and therefore results in an increased amount of free electrons, which survive to lower redshifts and affect the CMB anisotropies. This has been analyzed with increasing accuracy in a string of papers~\cite{Galli:2009zc,Slatyer:2009yq,Huetsi:2009ex,Cirelli:2009bb,Finkbeiner:2011dx,CMB_seealso}. Detailed constraints have been recently derived in~\cite{Hutsi:2011vx,Galli:2011rz,Natarajan:2012ry}, based on the most recent CMB data from WMAP and other observatories. The constraints are somewhat sensitive to the dominant DM annihilation channel: annihilation modes for which a portion of the energy is carried away by neutrinos or stored in protons have a lesser impact on the CMB; on the contrary the annihilation mode which produces directly $e^+e^-$ is the most effective one. In absolute terms, these constraints are among the most stringent ones on the DM annihilation cross section (see fig.~\ref{fig:cosmology}): they exclude the DM interpretation of the charged CRs excesses and they can rule out thermal DM with masses below 10 GeV or so.

\end{itemize}

%%%%%%%%%%%%%%%%%%%%%%%%%%%%%%%%%%%%%%%%%%%%%%%
\section{Uncertainties}
\label{uncertainties}

By definition, DM ID deals with features embedded in astrophysical signals, and therefore is `contaminated' by the uncertainties that affect the latter ones. In addition, there are specific uncertainties related to the DM itself and its particle physics origin. A list of the most important sources of uncertainties affecting DM Indirect Detection includes: 
\begin{itemize}
\item[$\Diamond$] The {\bf DM distribution} in the Galaxy and beyond, which articulates in a set of inter-related points: 
\begin{itemize}
\item[$\diamond$] What is the profile of DM distribution in the Milky Way? As briefly mentioned in Sec.~\ref{chargedscent}, adopting a peaked versus a cored profiled may make a difference of orders of magnitude in the prediction of the signals from DM, especially for those coming from the GC.
\item[$\diamond$] How reliable are the predictions we obtain from DM simulations? Numerical simulations can resolve DM halos only down to a certain mass and down to a certain spatial scale. Beyond that, we rely on extrapolations, sometimes of many orders of magnitude.
\item[$\diamond$] What is the effect of baryons on the DM profile? Baryons (i.e. ordinary matter making up clouds, stars, black holes and all the associated phenomena) start only now to be included in DM simulations. They could either steepen the DM profile at the center, thanks to adiabatic contraction, or smoothen it due to friction.
\item[$\diamond$] What is the local density of DM? Its value affects the normalization of the expected fluxes from the local environment but also that of the DM profiles in general.
\item[$\diamond$] What is the velocity distribution of DM particles in the galactic halo? The details of the velocity distribution of DM particles, including the escape velocity, mainly affect the nuclear recoil experiments of Direct Detection, but also some ID signatures, for instance the capture of DM particles in the Sun and thus the flux of neutrinos.   
\end{itemize} 
\item[$\Diamond$] The {\bf propagation} of charged CR. The propagation of $e^\pm$, $\bar p$, $\bar d\, $ in the Galaxy, sketched in Sec.~\ref{chargedscent}, contains a number of uncertain parameters: the diffusion coefficient (its normalization and energy dependence), the energy loss parameters, the values of the convective (and re-acceleration) velocity, the thickness of the diffusion box, the size of the effect of approaching the solar sphere... The fact that we cannot precisely predict them traces back to our fundamental ignorance about the details of the magnetic field inhomogeneities in the Galaxy, of the gas distribution, of background light, of solar-generated fields etc. We can determine reasonable ranges of values for these parameters, fitting a number of {\em ordinary} cosmic ray data, but still DM predictions can differ by more than one order of magnitude when the parameters are varied within these ranges. Progress in better understanding CR propagation will undoubtedly bring much progress in the DM field as well.

\item[$\Diamond$] {\bf Particle physics} uncertainties. This refers of course not to the unknown particle properties of DM such as the annihilation cross section or channels, which are the parameters to discover, but to the intervening tools from particle physics that are used to predict DM signatures. 
\begin{itemize}
\item[$\diamond$] For instance, almost every DM indirect search analysis uses the collider MonteCarlo code {\sc Pythia}~\cite{Sjostrand:2007gs} to compute the annihilation spectra, despite the fact that other codes are available and that in any case all codes have been designed and calibrated for the collider environment and in an energy range which (until recently) was much lower than the multi-TeV one of interest for some DM models. In Ref.~\cite{Cirelli:2010xx} the uncertainty related to MC issues is tentatively quoted at $\pm20\%$, although bigger surprises are possible for some channels. 
\item[$\diamond$] Some initially overlooked particle physics effects may turn out to be important, such as the EW corrections mentioned in Sec.~\ref{chargedscent}.
\item[$\diamond$] Some signals depend on poorly known nuclear or particle physics aspects, like e.g. the coalescence process for antideuterons. 
\end{itemize}

\item[$\Diamond$] The astrophysical {\bf `background'}. While technically not an uncertainty affecting the predictions from DM, the astrophysical background is arguably the source of most of the uncertainty affecting the interpretation of the ID signals in terms of DM. In every channel, knowing what can be ascribed to common (or even peculiar, but ordinary) astrophysics is the most important ingredient to be able to single out the exotic contribution. There is obviously no straightforward solution, but (i) to continuosly improve our knowledge of such backgrounds and (ii) to break the degeneracy requiring a genuine signal to be multimessenger, i.e. sticking out on top of at least two (hopefully unrelated) `backgrounds'.

\end{itemize}

%%%%%%%%%%%%%%%%%%%%%%%%%%%%%%%%%%%%%%%%%%%%%%%
\section{Cursory overview of the theory directions}
\label{theory}

\subsection{DM that wants to explain the charged CR excesses}

Inspired by the flurry of data discussed above, the field of DM model building has experienced a huge surge of activity in the latest few years. This is essentially because the DM properties individuated by the charged CR signals, and the associated constraints, pose a tough {\bf challenge for the traditional DM candidates}. Let us take the supersymmetric neutralino as a strawman and see how it fares as a candidate for the DM phenomenology outlined in page~\pageref{properties}: (i) The data require a multi-TeV DM mass, but this fits rather uncomfortably in a theory, such as SuSy, whose typical scale should be close to the mass of the higgs boson, if naturalness is a valid criterion. (ii) The data require leptophilicness, but typical neutralino annihilation channels do not distinguish between leptons and quarks, and often couple to gauge bosons, so a dangerous component of hadrons is generically expected. (iii) The data require a huge annihilation cross section into relatively light final states such as leptons, but the neutralino is a Majorana particle and, as such, its ($s$-wave) annihilation cross section is helicity suppressed by a large factor of $(m_f/m_{\mbox{\tiny DM}})^2$, where $m_f$ is the fermion mass. The scorecard for the neutralino does not look great. 
Of course these generic arguments can be circumvented in specific situations. Indeed several works have argued that it {\em is} possible to explain the CR excesses insisting on an MSSM DM candidate~\cite{Grajek:2008pg,Kane:2009if,Hooper:2008kv,Kadota:2010xm,Cotta:2010ej}. This, however, happens at the price of finely tuning at least some aspect: the properties of the $\bar p$ background, the position of a nearby DM clump, assuming (uplifted) resonances, finding just few configurations in a scatter plot, explaining only the positron rise and leaving the rest to ad-hoc astrophysics... . These examples are therefore perceived as somewhat anecdotical. Similar arguments would apply for other `traditional' DM frameworks, e.g. Kaluza-Klein (extradimensional) DM.

This is why the community has preferred to explore new model building possibilities and a lot of works have been published since 2008~\cite{PAMELAfrenzy}. Before sketching the main directions of this activity, let us first address an aspect which finds application in several models: how to obtain the large flux. 

\subsubsection{Tools for enhancements}
\label{enhancements}
So the question is: is it possible to reconcile the very large value of the annihilation rate (corresponding to $\langle \sigma v \rangle \gtrsim 10^{-23}\, {\rm cm}^3/{\rm sec}$) required `today' to fit the CR excesses and the smaller value (corresponding to $\langle \sigma v \rangle \simeq 3 \, 10^{-26}\, {\rm cm}^3/{\rm sec}$) individuated by the paradigm of DM production as a thermal relic in the Early Universe?
More generically, and thinking beyond the current CR excesses, a large flux would greatly increase our hopes for detection: are there ways to naturally obtain large rates?

The ingenuity of theorists has found at least three positive possible answers to these questions. Yes, it is possible:...
\begin{itemize}
\item[A)] Via an astrophysical {\bf boost factor}: the presence, in today's galactic halo, of DM overdensities predicted by numerical simulations boosts the annihilation rate (proportional to the squared density of DM particles). If the boost could reach a $\mathcal{O}(10^3)$ value, this would explain why the rate is much larger today, without modifying the cross section itself. The typical realistic values, however, have been proven to be ${\cal O}(10)$ at most~\cite{Lavalle:2006vb}. 

\item[B)] Annihilating via a {\bf resonance}~\cite{Cirelli:2008pk,Ibe:2008ye,Feldman:2008xs,An:2012uu}: if the resonance mass is just below twice the DM mass, the annihilation cross section becomes sensitive to the details of the velocity distribution of the DM particles; since (on average) DM particles are slower today than in the Early Universe, many more of them meet the conditions of resonant annihilation and therefore the rate is enhanced, provided that the relevant parameters are appropriately fine tuned. Typically one needs the mass of the resonance to differ by less than 1\% from twice the DM mass and the width of the resonance to be of the order of $10^{-5}$ of its mass, in order to obtain an enhancement of ${\cal O}(10^{3})$.  

\item[C)] Thanks to the {\bf Sommerfeld enhancement} (see~\cite{Sommerfeld, MDMastro, Cirelli:2008pk}, and then~\cite{Arkani,Sommerfeld2,Bovy:2009zs}), 
a non-perturbative effect which modifies the annihilation cross section in the regime of {\em small relative velocity} of the annihilating particles and in presence of an effectively {\em long-range force} between them. Indeed this well known quantum mechanical effect, first discussed by Sommerfeld in the context of positronium $e^+e^-$ annihilations under the effect of resummed $\gamma$ exchanges, can occur in DM annihilations if the two annihilating particles exchange an interaction mediated by a force carrier of mass $m_V$ and with a coupling constant $g \approx \sqrt{4 \pi \, \alpha}$ such that $\alpha M_{\rm DM}/m_V \gtrsim 1$. For very heavy DM particles ($\gtrsim 10$ TeV), the exchange of SM weak bosons can mediate the effect, in which case $\alpha$ is just the one of weak interactions and $m_V \approx m_{W^\pm,Z} \approx 100$ GeV. If a new force exist, however, mediated by a particle with mass $m_V \approx 1$ GeV and gauge-like coupling strength with DM particles only, even DM particles of mass $\lesssim$ 1 TeV would enjoy the Sommerfeld enhancement. The details of the enhancement of the cross--section are model dependent (see~\cite{Sommerfeld3,Sommerfeld4,Sommerfeld5} for some examples), but some general features can be identified: in particular, the enhancement shows an inverse proportionality with the relative velocity of the two particles, and it typically saturates to a maximum value when that is $\beta \lesssim 10^{-3}$.

\end{itemize}
While in some models the enhancement factors discussed above are natural or even unavoidable~\footnote{See e.g. the Sommerfeld enhancement in~\cite{MDMastro,MDMindirect}, published before the CR data made an enhancement needed, just to mention an example that I know well.}, it is fair to say that in many cases they have been used quite liberally in model building, along the lines of `If I miss a factor of $N$ to fit the data, I throw in A, B or C, or a combination thereof, until I get $N$'~\footnote{See e.g. the boost factor of $\sim$ 50 in~\cite{MDMindirect,MDMiDM}, introduced after the CR data showed that a little bit more of an enhancement was needed, just to mention an example that I know well.}.
Perhaps precisely for this reason, they have been studied in detail and they have become kind of a tool in the model builder toolbox. So they are likely to stay with us in the near future.

\subsubsection{Recent theory directions}
\label{theorydirections}

A possible categorization of the intense model building work of the past few years would identify the following classes:

\begin{itemize}
\item[$\bigstar$] {\bf Minimalistic} Dark Matter models. These are models loosely identified by the fact that they aim at providing a viable DM candidate insisting on introducing the minimal set of new particles beyond the Standard Model. They arised originally in opposition to the mainstream direction of obtaining DM as a byproduct of a more ambitious and comprehensive theory, such as SuSy or Extra Dimensions. The namesake Minimal Dark Matter (MDM)~\cite{Cirelli:2005uq} falls in this class, as well as less fundamentalist theories such as the model in~\cite{Mahbubani:2005pt}, the hidden vector~\cite{Hambye:2008bq}, the Inert Doublet Model (IDM)~\cite{Barbieri:2006dq,Hambye:2007vf} and others. 

Thanks to their relative simplicity, the models in this class are often free of many free parameter and therefore are quite predictive: their ID signatures can be computed and compared univocally with data. The MDM model, featuring, in its minimal realization, a 9.6 TeV DM particle annihilating almost exclusively into $W^+W^-$, initially excited its authors since it had predicted the size and shape of the positron rise (and the $\bar p$ null result) in PAMELA, provided that an astrophysical boost factor of $\sim$ 50 was adopted. Later it has been disfavored by the FERMI+HESS $e^\pm$ data, which prefer a lower mass, and $\gamma$-ray data, see e.g.~\cite{Pato:2009fn} (it remains a viable DM candidate --albeit obviously not an explanation for the CR excesses-- for a smaller --and, by the way, more realistic-- astrophysical boost factor). 
The phenomenology of the IDM shares a similar history~\cite{Nezri:2009jd}.

\item[$\bigstar$] Models with {\bf new dark forces} or, more generically, a rich Dark Sector. In this class fall most of the models whose construction has been directly stimulated by the CR excesses. 

The model which undoubtedly has most attracted attention and has best spelled out the ingredients is presented in~\cite{Arkani}, although similar ideas have been proposed before or around the same time~\cite{Pospelov,Feldman:2008xs,Cholis:2008vb,Nelson:2008hj,Cholis:2008qq,Nomura:2008ru,Bai:2008jt}. The model in~\cite{Arkani} features a TeV-ish DM particle which is sterile under the SM gauge group but which interacts with itself via a new force-carrying boson $\phi$ (with the strength of typical gauge couplings). The DM annihilation therefore proceeds through DM DM $\to \phi \phi$. A small mixing between $\phi$ and the electromagnetic current assures that $\phi$ eventually decays. Therefore the process of DM annihilation occurs in 2 steps: first two DM's go into two $\phi$'s and then each $\phi$'s, thanks to its mixing with a photon, goes into a couple of SM particles. The crucial ingredient is that the mass of $\phi$ is chosen to be light, of the order of $\lesssim$ 1 GeV. This simple assumption, remarkably, kills two birds with a stone. On one side, the exchange of $\phi$ realizes the Sommerfeld enhancement discussed in Sec.~\ref{enhancements}, thus providing a very large annihilation cross section today but preserving the thermal production of DM in the Early Universe. On the other side, $\phi$ can only decay into SM particles lighter than a GeV, i.e. electrons, muons and possibly pions, but not protons: this assures that the annihilation is leptophilic, for a simple kinematical reason. The model therefore fulfils all the requirements listed in page~\pageref{properties}. The construction can then be complicated {\em ad libitum}, e.g. assuming that the dark gauge group is non-abelian and the DM sits in a multiplet of such group, with small splitting between the components. This allows to accommodate other experimental anomalies, not discussed here.  

The kinematical argument is not the only one available to justify a leptophilic nature for DM. In the literature, variations have been proposed in which DM is coupled preferentially to leptons because it carries a lepton number~\cite{Phalen:2009xw}, because it shares a quantum number with a lepton~\cite{Cirelli:2008pk,Fox:2008kb}, because quarks live on another brane~\cite{Park:2009cs} or... `because I say so'~\cite{Harnik:2008uu}.\footnote{Strictly speaking, ref.s~\cite{Park:2009cs} and~\cite{Harnik:2008uu} should not fall in the class of models with new dark forces, since the former works in an extradimensional setup and the latter works in the framework of an effective field theory.}

The Indirect Detection phenomenology of these theories has of course been worked out in detail~\cite{Cholis:2008wq,Meade:2009iu,Mardon:2009rc}. In short, they can easily provide a fit to the charged CR excesses as good as, or better than, the one from ordinary annihilating DM, also thanks to the fact that the 2-step annihilation softens the spectrum of final $e^\pm$ and allows a better agreement with the data.~\footnote{See however~\cite{Serpico:2011wg} for some points of criticism.}
They are also subject to the same constraints from neutrinos and especially gamma rays: the bottom line of a series of analyses~\cite{Bergstrom:2008ag,Meade:2009rb,Papucci:2009gd,Cirelli:2010nh} is that, while the bounds are somewhat alleviated (essentially thanks to the smaller yield of $\gamma$-rays and the softening of the spectra mentioned above), there still remain a tension. 
The conclusion of, e.g.,~\cite{Finkbeiner:2010sm}, which also considers cosmological constraints, is that it is possible for these models to constitute an explanation of the CR excesses while remaining consistent with the constraints, albeit barely and at the condition of assuming a specific admixture of 4$e$ and 4$\mu$ final states and quite a specific mass for $\phi$.

\item[$\bigstar$] {\bf Decaying Dark Matter}. 
The possibility that Dark Matter consists of a particle that actually decays on a very long time scale has been considered since a long time, e.g. in the context of gravitino DM with R-parity violation~\cite{Buchmuller:2007ui}. 

More recently, this option has gained steam (see e.g.~\cite{Chen:2008dh,Yin:2008bs,Ishiwata:2008cv,Ibarra:2008jk,Chen:2008qs,Nardi:2008ix,Arvanitaki:2008hq}), precisely because of the cosmic ray excesses and the difficulty of explaining them with annihilating DM. Indeed, if the decay half-life is tuned to $\approx 10^{26}$ seconds (a figure possibly motivated by some high enegy physics scale suppressed operators)~\footnote{Of course, this value of $\tau_{\rm dec}$ is so much longer than the age of the Universe that the slow decay does not make a dent in the overall cosmological DM abundance and does not spoil the agreement with a number of astrophysical and cosmological observations, including the Cosmic Microwave Background~\cite{astroboundsdecayingDM}.} and if the production of hadrons is adequately suppressed by some a priori unrelated mechanism, the features needed to fit the data are obtained. A crucial advantage is also that gamma ray (and neutrino) constraints are in general less severe in these scenarios, as they are proportional to the first power of the DM density (and not the second, like for annihilations). 

The ID phenomenology of decaying DM has been studied in detail~\cite{Meade:2009iu,Ibarra:2008jk,Ibarra:2009tn}. Generally speaking, it is not much different from the phenomenology of annihilating DM, modulo a few points: i) decay channels other than the ordinary particle-antiparticle pair become possible for fermionic DM (e.g. DM $\to W^\pm \ell^\mp$, DM $\to \ell^+ \ell^- \nu$); ii) the local spectrum of charged CR is often somewhat harder for decaying DM than for annihilating DM, since in the latter case a sizable fraction of low energy particles manage to diffuse to the Earth from the high production regions in the inner Galaxy, softening the spectrum. All in all, it is possible to find fits to the data as good as, or even better than, those of ordinary annihilating DM.  
On the other hand, as anticipated, the gamma ray constraints from the Galaxy are not as stringent as in the case of annihilating DM. %For instance, even a cuspy profile such as Einasto is allowed. 
The most stringent bounds come from the isotropic $\gamma$ flux~\cite{Cirelli:2009dv,Papucci:2009gd,Cirelli:2012ut} and, recently, from galaxy clusters in FERMI data~\cite{Huang:2011xr,Zhang:2009ut}: in both cases decay half-lives shorter than $\sim$ few $\cdot 10^{26}$ sec or even up to $10^{27}$ sec (with the precise value depending on the $m_{\mbox{\tiny DM}}$ and the decay channel) are excluded~\footnote{The more stringent constraints from clusters derived in~\cite{Dugger:2010ys} are irrealistic.}. This represents a serious blow for the decaying DM interpretation of the CR excesses:~\cite{Cirelli:2012ut} shows that they are ruled out unless very conservative choices are adopted. 
The CMB constraints turn out to be not competitive for decaying DM~\cite{Chen:2003gz,CIPunpub}.

\end{itemize}

\subsection{DM that wants to explain the `130 GeV $\gamma$-ray line'}
\label{theorydirectionsline}

\vspace{-0.4cm}
\begin{figure}[!h]
\begin{minipage}{0.1\textwidth}
\begin{flushleft}
\includegraphics[width=\textwidth]{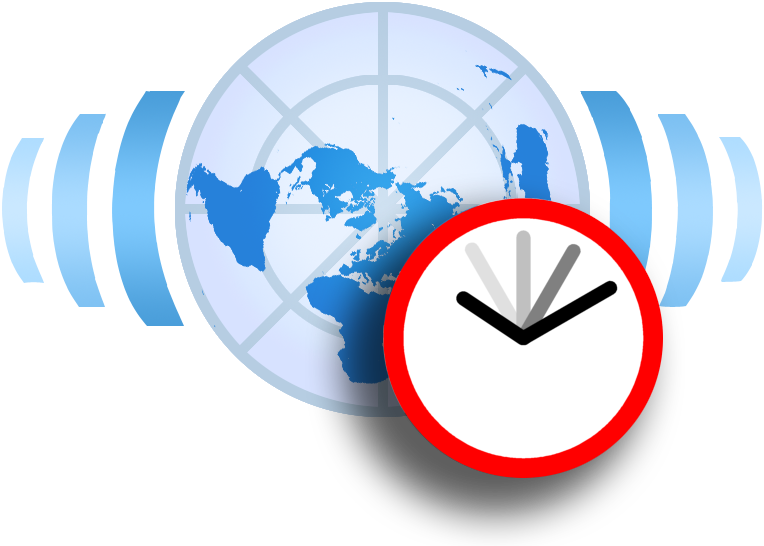} 
\end{flushleft}
\end{minipage}
\quad
\begin{minipage}{0.8\textwidth}
\begin{flushleft}
{\bf This section documents a current event.}\\ Information may change rapidly as the event progresses. {\footnotesize \it (August 2012)}
\end{flushleft}
\end{minipage}
\end{figure}
\vspace{-0.3cm}

The findings briefly introduced in Sec.~\ref{gammaclaims} have started inspiring some model building activity in the latest few months (for a more thorough review see~\cite{reviewBringmannWeniger}). The exercise is: produce a $\gamma\gamma$ line (or something narrow enough to look like a line) with relative cross section $\langle \sigma v \rangle_{\gamma\gamma} \sim 10^{-27}\, {\rm cm}^3/{\rm sec}$. The starting point is well known: since DM is electrically neutral, it does not directly couple to photons and therefore one needs some mediating mechanism in order to produce the $\gamma\gamma$ pair: 
(i) via a loop of charged particles, which can be SM ones~\cite{Jackson:2009kg}, SuperSymmetric ones~\cite{Kyae:2012vi,Acharya:2012dz,Das:2012ys} or from a new sector~\cite{Cline:2012nw,Park:2012xq,Weiner:2012gm}; 
(ib) via Chern-Simons terms~\cite{Dudas:2012pb}, the remnants of loops that have been integrated-away; 
(ii) via a normally subdominant DM-photon coupling, e.g. a magnetic dipole~\cite{Tulin:2012uq,Heo:2012dk,Cline2}; 
(iii) via axions~\cite{Lee:2012bq}\ldots \ 
As another option, one may produce a `top-hat' gamma ray spectrum by having DM annihilate into two metastable states $\phi$ that then decay into $\gamma\gamma$ (see e.g.~\cite{Ibarra:2012dw,Fan:2012gr}): the spectrum is flat between the edges $E_\pm = m_{\mbox{\tiny DM}}/2 \, \left(1 \pm \sqrt{1-m_\phi^2/m^2_{\mbox{\tiny DM}}} \right)$, so for $m_\phi$ close enough to $m_{\mbox{\tiny DM}}$ the top-hat fakes a line.
As yet another option one may interpret the feature at 130 GeV not as a proper line but rather as the peak of the internal bremsstrahlung discussed in~\ref{promptIB}~\cite{Bergstrom:2012bd,Shakya:2012fj}.

Next, one is confronted to two possible (possibly related) problems: the added `mediator' physics usually implies that the $\gamma\gamma$ process is suppressed with respect to other processes (for instance, the tree level annihilation into the particles running in the loops of (i)), but the needed $\langle \sigma v \rangle_{\gamma\gamma}$ required by FERMI data is relatively large and therefore the `other processes' are expected to have an even larger $\langle \sigma v \rangle$, with the result that: (1) they might be constrained by other observations (e.g. by FERMI data themselves, as discussed in Sec.~\ref{gammaclaims}); (2) they might tend to give a too large annihilation in the Early Universe and therefore an insufficient yield of the thermal DM abundance. This latter point is similar to the one already encountered for the charged CR anomalies (see Sec.~\ref{enhancements}), so that some of the tools already explored might apply. Additional proposed solutions include: (a) assuming that the tree level process is not accessible kinematically (e.g. the particles in the loop are heavier than the c.o.m. energy, i.e. the DM mass~\cite{Jackson:2009kg})~\footnote{Actually, in a model like~\cite{Jackson:2009kg}, one postulates that (or selects a scenario in which) DM annihilates (only) into top quarks via a $Z^\prime$. Demanding the correct relic abundance leads to requiring (a). Which naturally solves (1) and (2) and produces line-like features via (i).}, (b) engineering other mechanisms which obtain a correct relic abundance but which turn off today (e.g. coannihilations), (c) decoupling the relic DM from the one that produces signal, in model with multi-component DM, (d) giving up altogether the thermal relic mechanism\ldots \ See~\cite{Tulin:2012uq,Jackson:2013pjq} for lucid summaries.

Another quite generic `theory' consequence is that, wherever you can attach a $\gamma$ emission you can usually also attach a $Z$ or a $h$ emission, with the result that another line is expected at an energy $E = m_{\mbox{\tiny DM}} (1-m^2_{Z,h}/4m^2_{\mbox{\tiny DM}})$ (see e.g.~\cite{Rajaraman:2012db}). 

%%%%%%%%%%%%%%%%%%%%%%%%%%%%%%%%%%%%%%%%%%%%%%%
\section{Conclusions}
\label{conclusions}

Like possibly other fields in particle physics in these days, the one of Dark Matter Indirect Detection is currently characterized by an exciting mix of tantalizing hints, increasingly stringent constraints and ever rising hopes. 
%These are sometimes overlapping categories, acually (What looks like an unmistakable hint to some is a merciless constraint to others. In other words, businness as usual at the frontier of a rapidly advancing field.). 
According to my taste, and to the material presented above, I would classify:

\begin{itemize}
\item[$\vartriangle$] {\bf Hints}: the $e^+$ and $e^\pm$ excesses in PAMELA, FERMI and HESS. 

If interpreted in terms of DM annihilations, these point to a rather preposterous particle: multi-TeV, leptophilic and with a huge annihilation cross section. \\[2mm]
\phantom{{\bf Hints}:} the `130 GeV line' in FERMI data.

If interpreted in terms of DM annihilations, this seems to point to particle which reassures at first and then puzzles: its mass-scale and its spatial distribution are consistent with the phenomenologists' best dreams, but the large cross section is unsettling.

\item[$\triangledown$] {\bf Constraints}: the $\gamma$-ray measurements from FERMI and HESS (and, to a smaller extent, VERITAS and the other \v Cerenkov telescopes), the neutrino measurements from SuperKamiokande and ICECUBE and the CMB bounds based on WMAP. 

These rule out most or all of the parameter space for the preposterous DM above. They also put in a difficult spot some of the more exotic models discussed in Sec.~\ref{theorydirections}. 

On the other hand, they show how far-reaching these kinds of searches can be. They are starting to explore the parameter space of ordinary (thermally produced) DM and already rule out the lower mass portion of the tens-of-GeV range. 

\item[$\lozenge$] {\bf Hopes}: from the point of view of data, the $\gamma$-ray (and, probably to a smaller extent, the neutrino) telescopes will continue to play a major role: they will keep up their march across the parameter space and at some point they may see something totally convincing. The AMS-02 detector, onboard of the International Space Station, will probably deliver data in 2012: if the $e^\pm$ excesses are of astrophysical origin (as it is probably judicious to believe) then they constitute a formidable background to any future signal from DM in this channel; we have to place our hopes in another channel: the $\bar p$ one looks promising, and the $\bar d$ one (also explored by the GAPS experiment) may reserve surprises.

From the point of view of theory, the single most striking result of the recent activity is that the community has diversified its interests and explored many new directions, both in terms of phenomena that might enhance the DM signal (see Sec.~\ref{enhancements}) and in terms of model-building (see Sec.~\ref{theorydirections} and~\ref{theorydirectionsline}). The hope is that one or more of these interesting ramifications will bear fruit soon.
\end{itemize}

\acknowledgments
I thank all my collaborators for the work performed together and for useful discussions, in particular Paolo Panci, Alessandro Strumia, Gabrijela Zaharijas and above all Pasquale Serpico. 
My work is supported in part by the French national research agency ANR under contract ANR 2010 BLANC 041301 and by the EU ITN network UNILHC. I gratefully acknowledge a grant from the Comit\'e Fran\c cais de Physique (CFP) which financed in part my participation to the 2011 Lepton-Photon Symposium and I would like to thank the organizers of such Symposium for the very enjoyable atmosphere.

%\bibliographystyle{pramana}
%\bibliography{references}

\end{document}